\documentclass[11pt]{article}
\usepackage{graphicx}
\usepackage[utf8]{inputenc} 
\pdfoutput=1

\usepackage{amsmath,amsfonts,amssymb,epsfig}
\usepackage{slashed}
\numberwithin{equation}{section}

\usepackage{graphicx}
\usepackage{cancel}
\usepackage{cite}
\usepackage{color}

\usepackage{xspace}

\newcommand{\SARAH}{{\tt SARAH}\xspace}

\newcommand{\exclude}[1]{}
\def\nn{\nonumber}

\def\beq{\begin{equation}}
\def\eeq{\end{equation}}
\def\bal{\begin{align}}
\def\eal{\end{align}}

\def\s2b{s_{2\beta}}
\def\c2b{c_{2\beta}}

\long\def\symbolfootnote[#1]#2{\begingroup%
\def\thefootnote{\fnsymbol{footnote}}\footnote[#1]{#2}\endgroup}

\def\2b2[#1,#2][#3,#4]{\left( \begin{array}{cc} #1 & #2 \\ #3 & #4 \end{array}
\right)}
\def\3b3[#1,#2,#3][#4,#5,#6][#7,#8,#9]{\left( \begin{array}{ccc} #1 & #2 &#3 \\
#4 & #5 & #6\\#7&#8&#9\end{array} \right)}
\def\thv[#1,#2,#3]{\left( \begin{array}{c} #1 \\ #2 \\ #3 \end{array} \right)}
\def\twv[#1,#2]{\left( \begin{array}{c} #1 \\ #2 \end{array} \right)}

\def\twomat[#1,#2][#3,#4]{\left( \begin{array}{cc} #1 & #2 \\ #3 & #4 \end{array} \right)}
\def\threemat[#1,#2,#3][#4,#5,#6][#7,#8,#9]{\left( \begin{array}{ccc} #1 & #2 & #3\\ #4 & #5 & #6 \\ #7 & #8 & #9 \end{array} \right)}
\def\twovec[#1,#2]{\left( \begin{array}{c} #1  \\ #2 \end{array} \right)}

\def\C{\mathcal}

\def\smallMSSM{{\rm{\scriptscriptstyle MSSM}}}

\def\smallZ{{\scriptscriptstyle Z}}

\def\MS{M_S}
\def\MZ{m_\smallZ}

\def\msbar{{\ov {\rm MS}}}
\def\beq{\begin{equation}}
\def\eeq{\end{equation}}
\def\bea{\begin{eqnarray}}
\def\eea{\end{eqnarray}}

\def\sq2{\sqrt{2}}
\def\drbar{{\ensuremath{ \overline{\rm DR}}}}
\def\msbar{\overline{\rm MS}}
\def\smalldrbar{\scriptscriptstyle{\overline{\rm DR}}}

\def\smallOS{{\scriptscriptstyle {\rm OS}}}

\def\tb{\tan\beta}
\def\gl{\tilde{g}}
\def\mg{m_{\gl}}
\def\g{\mg^2}

\def\mgi{m_{\tilde{g}_i}}
\def\gi{\mgi^2}
\def\os{m_{O_1}^2}

\def\oi{m_{O_i}^2}

\def\mix{X}
\def\mixt{\widetilde X}

\def\msusy{\MS}

\def\at{\alpha_t}
\def\ab{\alpha_b}
\def\as{\alpha_s}
\def\oas{{\cal O}(\as)}

\def\oatas{{\cal O}(\at\as)}
\def\oabas{{\cal O}(\ab\as)}

\def\mt{m_t}

\def\t{\mt^2}

\def\tu{m_{\tilde{t}_1}^2}
\def\td{m_{\tilde{t}_2}^2}
\def\tul{m_{\tilde{t}_1}}
\def\tdl{m_{\tilde{t}_2}}
\def\ti{m_{\tilde{t}_i}^2}

\def\sdt{s_{2\theta_t}}
\def\cdt{c_{2\theta_t}}
\def\cdtt{c_{2\bar\theta_t}}
\def\sdtt{s_{2\bar\theta_t}}
\def\ttbar{\bar \theta_{\tilde t}}

\def\DVt{\frac{\partial \Delta V}{\partial \mt^2 }}
\def\DVtu{\frac{\partial \Delta V}{\partial \tu}}
\def\DVtd{\frac{\partial \Delta V}{\partial \td}}
\def\DVcdtq{\frac{\partial \Delta V}{\partial \cdtt^2}}

\def\DVtt{\frac{\partial^{\,2} \Delta V}{(\partial \mt^2)^2 }}
\def\DVttu{\frac{\partial^{\,2} \Delta V}{\partial \mt^2 \partial \tu }}
\def\DVttd{\frac{\partial^{\,2} \Delta V}{\partial \mt^2 \partial \td }}
\def\DVtutu{\frac{\partial^{\,2} \Delta V}{(\partial \tu)^2 }}
\def\DVtutd{\frac{\partial^{\,2} \Delta V}{\partial \tu \partial \td }}
\def\DVtdtd{\frac{\partial^{\,2} \Delta V}{(\partial \td)^2}}
\def\DVcdtqtu{\frac{\partial^{\,2} \Delta V}{\partial \cdtt^2 \partial \tu }}
\def\DVcdtqtd{\frac{\partial^{\,2} \Delta V}{\partial \cdtt^2 \partial \td }}
\def\DVcdtqt{\frac{\partial^{\,2} \Delta V}{\partial \cdtt^2 \partial \mt^2 }}
\def\DVcdtqcdtq{\frac{\partial^{\,2} \Delta V}{(\partial \cdtt^2)^2 }}

\begin{document}

\begin{titlepage}

\begin{flushright}
\end{flushright}
\begin{center}

\vspace{1cm}

{\LARGE \bf Leading two-loop corrections to the Higgs boson masses} 
\vskip 0.3cm
{\LARGE \bf in SUSY models with Dirac gauginos}

\vspace{1cm}

{\Large Johannes~Braathen,$^{\!\!\!\,a,b}$~ 
Mark~D.~Goodsell$^{\,a,b}$~ and Pietro~Slavich$^{\,a,b}$}

\vspace*{5mm}

{\sl ${}^a$ LPTHE, UPMC Univ.~Paris 06,
  Sorbonne Universit\'es, 4 Place Jussieu, F-75252 Paris, France}
\vspace*{2mm}\\
{\sl ${}^b$ LPTHE, CNRS, 4 Place Jussieu, F-75252 Paris, France }
\end{center}
\symbolfootnote[0]{{\tt e-mail:}}
\symbolfootnote[0]{{\tt braathen@lpthe.jussieu.fr}}
\symbolfootnote[0]{{\tt goodsell@lpthe.jussieu.fr}}
\symbolfootnote[0]{{\tt slavich@lpthe.jussieu.fr}}

\vspace{0.7cm}

\abstract{We compute the two-loop $\oatas$ corrections to the Higgs
  boson masses in supersymmetric extensions of the Standard Model with
  Dirac gaugino masses. We rely on the effective-potential technique,
  allow for both Dirac and Majorana mass terms for the gluinos, and
  compute the corrections in both the $\drbar$ and on-shell
  renormalisation schemes. We give detailed results for the MDGSSM and
  the MRSSM, and simple approximate formulae valid in the decoupling
  limit for all currently-studied variants of supersymmetric models
  with Dirac gluinos. These results represent the first explicit
  two-loop calculation of Higgs boson masses in supersymmetric models
  beyond the MSSM and the NMSSM.}

\vfill

\end{titlepage}

\tableofcontents

\setcounter{footnote}{0}

\section{Introduction}
\label{sec:intro}

In anticipation of new results from the run II of the LHC,
supersymmetry (SUSY) as a framework remains the leading candidate for
physics beyond the Standard Model (SM). However, the discovery of a
SM-like Higgs boson with relatively large mass and the lack of
observation of coloured superparticles have spurred considerable
interest in SUSY realisations beyond the Minimal Supersymmetric
Standard Model (MSSM). A notable extension beyond the minimal case is
to allow Dirac masses for the gauginos
\cite{Fayet:1978qc,Polchinski:1982an,Hall:1990hq,Fox:2002bu,Nelson:2002ca,Antoniadis:2006uj},
in particular instead of -- but possibly in addition to -- Majorana
ones. Among the reasons for the growing interest in this scenario are
that Dirac gaugino masses relax constraints on squark masses (through
suppressing production)~\cite{Heikinheimo:2011fk, Kribs:2012gx,
  Kribs:2013oda} and flavour constraints~\cite{Kribs:2007ac,
  Fok:2010vk, Dudas:2013gga}, and that they increase the naturalness
of the model (because the operators are
\emph{supersoft}~\cite{Fox:2002bu} and the SM-like Higgs boson mass is
enhanced at tree level~\cite{Belanger:2009wf, Benakli:2011kz}).

Dirac gaugino masses require the addition of two fermionic degrees of
freedom (i.e.,~an extra Weyl spinor) for each gaugino. We can then
write a mass term that respects a global chiral symmetry, which in
SUSY models is promoted to a global $U(1)$ $R$-symmetry. We also
require the same number of extra scalar degrees of freedom as
fermionic ones; this implies that after electroweak symmetry breaking
(EWSB) we have four new neutral scalar degrees of freedom compared to
the MSSM, which may mix with the neutral scalars of the Higgs
sector. The new states are packaged in an adjoint chiral multiplet for
each gauge group, which should also have couplings to the Higgs
scalars, possibly enhancing the SM-like Higgs boson mass at both tree
and loop level.

There is more than one way to construct a Dirac-gaugino extension of
the MSSM. The minimal choice, which we will denote as the Minimal
Dirac Gaugino Supersymmetric Standard Model (MDGSSM), consists in
simply adding the adjoint chiral multiplets to the field content of
the MSSM, and allowing for all gauge-invariant terms in the
superpotential and in the soft SUSY-breaking Lagrangian.
The reader should note that in recent works~\cite{Benakli:2014cia,
  Goodsell:2015ura, Benakli:2016ybe} the term MDGSSM has also been
used to describe a unified scenario where extra lepton-like states are
added to ensure natural gauge-coupling unification, but the
distinction will be irrelevant for this paper.

If we wish to avoid large Majorana masses for the gauginos -- and
benefit from simpler SUSY-breaking scenarios~\cite{Nelson:1993nf} --
we should avoid $R$-symmetry breaking in the soft terms, which also
means removing the MSSM-like $A$-terms; this then naturally embeds
into gauge-mediated scenarios~\cite{Antoniadis:2006uj, Amigo:2008rc,
  Benakli:2008pg, Benakli:2010gi, Carpenter:2010as, Abel:2011dc,
  Csaki:2013fla}. We may retain a $B_\mu$ term since it is required
for EWSB and will not generate Majorana masses through renormalisation
group evolution. The variant without a $\mu$ term is called
$\slashed{\mu}$SSM or $\mu$-less MSSM~\cite{Nelson:2002ca}, and can be
considered a special case of this model (note that, as studied in
ref.~\cite{Benakli:2012cy}, the $\slashed{\mu}$SSM is currently
challenged by electroweak precision measurements).

On the other hand, if we choose to retain the $R$-symmetry as exact
(possibly broken only by gravitational effects) then one popular
construction is the Minimal $R$-symmetric Supersymmetric Standard
Model, or MRSSM~\cite{Kribs:2007ac}: two additional Higgs-like
superfields are included, which couple in the superpotential to the
regular Higgs doublets but obtain no expectation value. They allow the
Higgs fields $H_u$ and $H_d$ to both have zero $R$-charge and
contribute to EWSB without violating the $R$-symmetry. An even more
minimal realisation is the MMRSSM~\cite{Frugiuele:2011mh,
  Bertuzzo:2014bwa}, where the down-type Higgs $H_d$ and its
$R$-partner are missing, a sneutrino then playing the role of
$H_d$. Another option to preserve $R$-symmetry is the supersymmetric
one-Higgs-doublet model~\cite{Davies:2011mp}: starting from the field
content of the MDGSSM, the singlet adjoint superfield is missing and
the down-type Higgs does not develop an expectation value, therefore
the bino is massless up to anomaly-mediation contributions.

The extended Higgs sectors of these theories have an interesting and
varied phenomenology. From past experience in the study of the Higgs
sector of the MSSM and of the Next-to-Minimal Supersymmetric Standard
Model (NMSSM), we expect the radiative corrections to the Higgs boson
masses in Dirac-gaugino models to be crucial to obtain a reasonable
precision and rule in/out scenarios, or assess their naturalness.
For the MSSM, the corrections to the Higgs boson masses have been
computed at two loops in the limit of vanishing external
momenta~\cite{Hempfling:1993qq, Heinemeyer:1998jw, Heinemeyer:1998kz,
  Zhang:1998bm, Heinemeyer:1998np, Espinosa:1999zm, Espinosa:2000df,
  Degrassi:2001yf, Brignole:2001jy, Brignole:2002bz, Martin:2002iu,
  Martin:2002wn, Dedes:2003km, Heinemeyer:2004xw}, and the dominant
momentum-dependent two-loop corrections~\cite{Martin:2004kr,
  Borowka:2014wla, Degrassi:2014pfa} as well as the dominant
three-loop corrections~\cite{Martin:2007pg, Harlander:2008ju,
  Kant:2010tf} have also been obtained.\footnote{We focused here on
  genuine two- and three-loop corrections in the MSSM with real
  parameters, but significant efforts have also been devoted to
  Higgs-mass calculations in the presence of CP-violating phases, and
  to the computation of higher-order corrections via
  renormalisation-group techniques.} For the NMSSM, beyond the
one-loop level only the two-loop corrections involving the strong
gauge coupling together with the top or bottom Yukawa couplings,
usually denoted as $\oatas$ and $\oabas$, have been
computed~\cite{Degrassi:2009yq, Muhlleitner:2014vsa}. In contrast, in
other supersymmetric extensions of the SM there have been no
\emph{explicit} calculations of the Higgs masses beyond one-loop
results.

On the other hand, the public tool \SARAH~\cite{Staub:2008uz,
  Staub:2009bi, Staub:2010jh, Staub:2012pb, Staub:2013tta,
  Staub:2015kfa} can, for a generic supersymmetric model,
automatically compute the full one-loop corrections to all particle
masses, as well as the two-loop corrections to the neutral-scalar
masses in the limit of vanishing electroweak gauge couplings and
external momenta~\cite{Goodsell:2014bna,Goodsell:2015ira},
implementing and extending the general two-loop results of
refs.~\cite{Martin:2001vx, Martin:2003it}.
Recently, \SARAH~has made it possible to analyse at the two-loop level
the Higgs sector of several non-minimal extensions of the
MSSM, see refs.~\cite{Goodsell:2014pla, Dreiner:2014lqa, Nickel:2015dna,
  Goodsell:2015yca, Goodsell:2016udb}.  Of particular relevance for
this work, it has allowed for Dirac-gaugino masses since version {\tt
  3.2}~\cite{Staub:2012pb}, incorporating also the results of
ref.~\cite{Goodsell:2012fm}.
Indeed, \SARAH has been used for detailed phenomenological analyses of
the MDGSSM at one loop in ref.~\cite{Benakli:2012cy} and at two loops
in refs.~\cite{Benakli:2014cia, Goodsell:2015ura}; and also for the
MRSSM at one loop in ref.~\cite{Diessner:2014ksa} and two loops in
refs.~\cite{Diessner:2015yna, Diessner:2015iln}.

However, while such a numerical tool for generic models fulfils a
significant need of the community, it is also important to have
explicit results for specific models, and not just as a
cross-check. In this work we shall compute the leading $\oatas$
corrections to the neutral Higgs boson masses in both the MDGSSM and
MRSSM, relying on the effective-potential techniques developed in
ref.~\cite{Degrassi:2001yf} for the MSSM and in
ref.~\cite{Degrassi:2009yq} for the NMSSM. This has the following
advantages:

\begin{itemize}
\item We compute the $\oatas$ corrections in both the $\drbar$ and
  on-shell (OS) renormalisation schemes. The latter turns out to be
  particularly useful in scenarios with heavy gluinos -- a feature of
  many Dirac-gaugino models in the literature -- where the use of
  $\drbar$ formulae for the two-loop Higgs-mass corrections can lead
  to large theoretical uncertainties.

\item We have written a simple and fast stand-alone code implementing
  our results, which we make available upon request (indeed, a version
  of the code is already included in \SARAH).

\item We use our results to derive simple approximate expressions for
  the most important two-loop corrections, applicable in any Dirac
  gaugino model.

\end{itemize}

The outline of the paper is as follows. In
section~\ref{sec:definition} we define the important parameters of our
theory. In section~\ref{sec:2loopcalc} we present our results for the
general case, the MDGSSM and the MRSSM, show how we compute the shift
to the OS scheme, and give simplified formulae for the SM-like Higgs
boson mass either for a common SUSY-breaking scale or for a heavy
Dirac gluino. In section~\ref{sec:numerical} we give numerical
examples of our results, illustrating the advantages of our approach
and also discussing the inherent theoretical uncertainties. We
conclude in section~\ref{sec:conclusion}. Explicit expressions for the
derivatives of the effective potential are given in an appendix.


\section{Definition of the theory}
\label{sec:definition}

\subsection{Adjoint multiplets and the supersoft operator}
\label{sec:adjoints}

In order to give gauginos a Dirac mass it is necessary to pair each
Weyl fermion of the vector multiplets with another Weyl fermion
$\chi_\Sigma$ in the adjoint representation of the gauge group. These
adjoint fermions sit inside chiral superfields, which we shall denote
collectively
$\mathbf{\Sigma}^a \,=\, \Sigma^a \,+\, \sqrt{2}\, \theta
\chi^a_\Sigma \,+ \ldots\,$, where the lowest-order component
$\Sigma^a$ is a complex scalar. In models with softly-broken
supersymmetry\footnote{It has also been suggested, e.g.~in
  ref.~\cite{Martin:2015eca}, that Dirac masses could arise through
  other operators; we do not consider them as they potentially
  correspond to a hard breaking of SUSY.}, the Dirac gaugino mass
arises only in the \emph{supersoft} operator:
\bea
\C{L}_{\mathrm{supersoft}} &=& 
\int d^2 \theta \,\sqrt{2}\, m_D \,\theta^\alpha \,\mathbf{W}_\alpha^a\, 
\mathbf{\Sigma}^a 
~+~ {\rm h.c.}\nn\\
&\supset& 
- m_D \,\lambda^a \chi^a_\Sigma  ~+~ \sqrt{2} \,m_D \,\Sigma^a\, D^a\
~+~ {\rm h.c.}, 
\eea
where $\mathbf{W}_\alpha^a \,=\, \lambda^a_\alpha \,+\, \theta_\alpha D^a
\,+\ldots$ is the field-strength superfield. Integrating out the
auxiliary field $D^a$ leads to mass terms for the adjoint scalars, as
well as to trilinear interactions between the adjoint scalars and the
MSSM-like scalars, which we collectively denote as $\phi$:
\beq
\label{dterms}
\C{L} ~\supset~  
- ( m_D\,\Sigma^a + m_D^*\, \Sigma^{a\,*})^2 
- \sqrt{2} \,g\, ( m_D \,\Sigma^a +  m_D^*\, \Sigma^{a\,*})\, \phi^* 
\,t^a \,\phi~,
\eeq
where $t^a$ are the generators of the gauge group in the
representation appropriate to $\phi$, and a sum over the gauge indices
of $\phi$ is understood. Considering all sources of mass terms for the
adjoint scalars,
\beq
\label{adjointmass}
\C{L} ~\supset~
- (m_\Sigma^2 + 2\,|m_D|^2)\, \Sigma^{a\,*} \Sigma^a 
-\frac{1}{2}\,(B_\Sigma + 2\,m_D^2)\,\Sigma^a \,\Sigma^a 
-\frac{1}{2}\,(B_\Sigma^* + 2\, m_D^{*\,2})\, \Sigma^{a\,*}\Sigma^{a\,*}~,
\eeq
where $m_\Sigma^2$ includes in general contributions from both the
superpotential and the soft SUSY-breaking Lagrangian, and $B_\Sigma$ is a
soft SUSY-breaking bilinear term. In addition, mixing with the
MSSM-like Higgs scalars may be induced, upon EWSB, by the $D$-term
interactions in eq.~(\ref{dterms}), as well as by superpotential
interactions.

We shall denote the adjoint multiplet for $U(1)_Y$ as a singlet
$\mathbf{S} = S \,+\, \sqrt{2}\, \theta \chi_S\,+ \ldots\,$, the one
for $SU(2)_L$ as a triplet
$\mathbf{T}^a = T^a \,+\, \sqrt{2}\, \theta \chi_T^a\,+ \ldots\,$, and
the one for $SU(3)$ as an octet
$\mathbf{O}^a = O^a + \sqrt{2}\, \theta \chi_O^a\,+ \ldots\,$. In this
paper we shall be interested only in the two-loop corrections to the
Higgs masses involving the strong gauge coupling $g_s$, thus the
relevant trilinear couplings in eq.~(\ref{dterms}) will be the ones
involving the octet scalar and the squarks. 

We shall make the additional restriction that the octet scalar only
interacts via the strong gauge coupling and the above trilinear terms,
equivalent to the assumption that it has no superpotential couplings
or soft trilinear couplings other than with itself. This shall
simplify the computations, and it is true for \emph{almost} all
variants of Dirac gaugino models studied so far. To have
renormalisable Yukawa couplings between the octet and the MSSM fields
we would need to add new coloured states (such as a vector-like
top). However, in the most general version of the MDGSSM there could
also be terms that violate the above assumption -- which have only
recently attracted attention~\cite{Cohen:2016kuf, Benakli:2016ybe} --
namely couplings between the singlet and the octet of the form
\beq 
 W ~\supset~ \frac{1}{2}\, \lambda_{SO} \,\mathbf{S}
\,\mathbf{O}^a \mathbf{O}^a, \qquad
\mathcal{L} ~\supset ~-\frac{1}{2} \,T_{SO} \,S \,O^a O^a \,+\, {\rm
  h.c.}~.
\label{EQ:SOO}
\eeq
The coupling $\lambda_{SO}$ is typically neglected because it violates
$R$-symmetry and leads to Majorana gaugino masses: for example, in the
restricted version of the MDGSSM or the $\slashed{\mu}SSM$ the
$R$-symmetry violation is assumed to only occur in the Higgs sector
and possibly only via gravitational effects. On the other hand, there
is no symmetry preventing the generation of $T_{SO}$, but it is
typically difficult for it to obtain a phenomenologically significant
magnitude, hence it has been neglected -- see \cite{Benakli:2016ybe}
for a full discussion (and for cases when it could be
large). Furthermore, $T_{SO}$ is irrelevant in the
decoupling limit (when the singlet $S$ is heavy) that we shall employ
later in our simplified formulae.

With the above assumptions, we can make a rotation of the superfield
$\mathbf{O}^a$ such that we can take $m_D$ to be real without loss of
generality, but we cannot simultaneously require that the soft
SUSY-breaking bilinear $B_O$ be real without additionally imposing CP
invariance.  The octet mass terms are then
\beq
\label{octetmass}
\C{L} ~\supset~ 
- m_O^2 \, O^{a\,*} O^a 
~-~ \frac{1}{2} \,B_O\, O^a \,O^a 
~-~ \frac{1}{2} \,B_O^*\, O^{a\,*} O^{a\,*} - m_D^2 \,( O^a + O^{a\,*})^2~.
\eeq
If $B_O$ is not real, the real and imaginary parts of the octet scalar
mix with each other.  Their mass matrix can be diagonalised with a
rotation by an angle $\phi_O\,$,
\beq
O^a ~=~ \frac{e^{i\phi_O}}{\sqrt{2}} \,(O_1^a + i \,O_2^a)~,~~~~~~~~
\phi_O ~=~ -\frac{1}{2}\,{\rm Arg}\left(B_O + 2\,m_D^2\right)~,
\eeq
to obtain the two mass eigenvalues
\beq
m^2_{O_{1,2}} ~=~
m_O^2 + 2 \,m_D^2 \pm |2 \,m_D^2 + B_O|~.
\eeq
Then the trilinear couplings of the octet mass eigenstates $O^a_{1,2}$
to squarks $\tilde q_L$ and $\tilde q_R$ read
\beq
\label{eq:interaction}
\C{L} ~\supset~ 
-2 \, g_s\, m_D \,( \cos \phi_O \, O_1^a - \sin \phi_O \, O_2^a)~ 
(\tilde q_L^*\, t^a\, \tilde q_L - \tilde q_R^*\, t^a\, \tilde q_R)~,
\eeq
where $t^a$ are the generators of the fundamental representation of
$SU(3)$.  These couplings lead to new (compared to MSSM and NMSSM)
contributions to the two-loop effective potential involving the octet
scalars which will affect the Higgs masses. We remark that, since in
eq.~(\ref{octetmass}) the superpotential mass term $m_D^2$ affects
only the real part of the octet scalar, the mixing angle $\phi_O$ is
suppressed by $m_D^2$ in the limit where the latter is much larger
than the soft SUSY-breaking mass terms. In particular,
\beq
\label{limitphiO}
\cos\phi_O~\approx~ 1
~+~ {\cal O}\left(m_D^{-4}\right)~,
~~~~~~~~~~
\sin\phi_O~\approx~-\frac{{\rm Im}(B_O)}{4\,m_D^2}
~+~ {\cal O}\left(m_D^{-4}\right)~.
\eeq
For the remainder of this paper, we shall restrict our attention to
the CP-conserving case. This is motivated by clarity and simplicity in
the calculations, and also physically in that there are strong
constraints upon CP violation, even in the Higgs
sector\cite{Ibrahim:1997gj, Ibrahim:1998je, Abel:2001vy,
  Pospelov:2005pr}. However, we shall make an exception in allowing a
non-zero angle $\phi_O$, because it is particularly simple to do so,
and its effects are only felt at an order beyond that considered here:
it generates CP-violating phases in the stop mass matrix at two loops,
and in the Higgs mass at three. This is because the couplings in
eq.~(\ref{eq:interaction}) are real, and phases only appear in the
octet scalar-gluino-gluino vertex.

\subsection{Gluino masses and couplings}
\label{sec:gauginos}

In the case of Dirac gauginos, there is mixing between the Weyl
fermion of the gauge multiplet $\lambda^a$ and its Dirac partner
$\chi^a_\Sigma$. We shall allow in general both Majorana and Dirac
masses which, in two-component notation, we write as
\beq
\label{gauginomass}
\mathcal{L}~\supset~
- \,\frac{1}{2}\, M_\lambda \,\lambda^a\lambda^a 
\,-\,\frac{1}{2}\, M_\Sigma\,\chi^a_\Sigma\chi^a_\Sigma 
\,-\, m_D\,\lambda^a\chi^a_\Sigma
\,+\,{\rm h.c.}~.
\eeq
As mentioned in the previous section, we can define $m_D$ to be
real. In general we cannot remove the phases from both $M_\lambda$ and
$M_\Sigma$; however, as also mentioned above, we shall not consider CP
violation in the gluino sector, and thus take all three masses to be
real. We then rotate $\lambda^a$ and $\chi^a_\Sigma$ to mass
eigenstates $\lambda^a_1$ and $\lambda^a_2$ via a mixing matrix
$R_{ij}$, so that
\beq
\lambda^a ~=~ R_{11} \, \lambda^a_1 \,+\, R_{12} \,\lambda^a_2\,,
 \qquad 
\chi^a_\Sigma ~=~ R_{21}\, \lambda^a_1 \,+\, R_{22}\,\lambda_2^a~.
\eeq
In four-component notation, this leads in general to two Majorana
gauginos with different masses. In case of a pure Dirac mass, however,
we obtain two Majorana gauginos with degenerate masses $|m_{\lambda_1}|
= |m_{\lambda_2}| = |m_D|\,$, which can also be combined in a single
Dirac gaugino.

We recall that in the models of interest in this paper there are no
Yukawa couplings of the additional octet superfield, therefore the two
gluino mass-eigenstates only couple to quarks and squarks via their
gaugino component $\lambda^a$. In particular, the couplings of each
(four-component) gluino $\tilde g^a_i$ are simply related to the
couplings of the usual (N)MSSM gluino by an insertion of the mixing
matrix:
\beq
\mathcal{L}~\supset~ - \sqrt{2}\,g_s\, R_{1i}\,\left[\,
\tilde q_L^*\,t^a\,(\overline{\tilde g^a_i}\,P_L\,q) 
\,-\, (\overline q \, P_L \,\tilde g^a_i)\, t^a\,\tilde q_R \,\right] 
~+~{\rm h.c.}~,
\eeq 
where a sum over the $SU(3)$ indices of quarks and squarks is again
understood. Consequently, as we shall see below, the gluino
contribution to the two-loop effective potential in Dirac-gaugino
models can be trivially recovered from the known results valid in the
MSSM and in the NMSSM.

\subsection{Higgs sector}

We now consider the Higgs sector of the theory. Dirac gaugino models
extend the (N)MSSM, so we shall assume that we have at least the usual
two Higgs doublets $H_u$ and $H_d$. To these we must add the adjoint
scalars $S$ and $T^a$ mentioned above, which mix with the Higgs
fields. The couplings of the adjoint scalars, as well as the presence
of any additional fields in the Higgs sector, will, however, depend on
the model under consideration. In the following we shall focus on the
minimal Dirac-gaugino extension of the MSSM, the MDGSSM, and on the
minimal $R$-symmetric extension, the MRSSM.

In the MDGSSM there are no additional superfields apart from the
adjoint ones, and the superpotential reads
\bea
W 
&=& W_{\mathrm{Yukawa}} \,+\, W_{\mathrm{MDGSSM}}~, \\[2mm]
\label{WYUK}
W_{\mathrm{Yukawa}} &=& 
Y_u^{ij} \,\mathbf{U}_i \mathbf{Q}_j\cdot\mathbf{H_u} 
\,-\, Y_d^{ij}\, \mathbf{D}_i \mathbf{Q}_j \cdot\mathbf{H_d} 
\,-\, Y_e^{ij}\, \mathbf{E}_i \mathbf{L}_j \cdot\mathbf{H_d}~, \\[2mm]
\label{WDGSSM}
W_{\mathrm{MDGSSM}} &=& 
(\mu + \lambda_{S} \,\mathbf{S} ) \,\mathbf{H_u}\cdot\mathbf{H_d} 
\,+\, \lambda_T\, \mathbf{H_d}\cdot \mathbf{T}^a\,\sigma^a\,\mathbf{H_u}
\,+\, W_{\Sigma}~,
\eea
where $\sigma^a$ are Pauli matrices, and the dot-product denotes the
antisymmetric contraction of the $SU(2)_L$ indices. In addition to the
terms explicitly shown in eqs.~(\ref{WYUK}) and (\ref{WDGSSM}), the
most general renormalisable superpotential contains terms involving
only the adjoint superfields -- namely, mass terms for each of them,
all trilinear terms allowed by the gauge symmetries, and a linear term
for the singlet -- which we denote collectively as $W_{\Sigma}$. The
most general soft SUSY-breaking Lagrangian for the MDGSSM contains
non-holomorphic mass terms for all of the scalars, as well as Majorana
mass terms for the gauginos, plus $A$-type (i.e., trilinear), $B$-type
(i.e., bilinear) and tadpole (i.e., linear) holomorphic terms for the
scalars with the same structure as the terms in the
superpotential. With the assumption, discussed in
section~\ref{sec:adjoints}, that we neglect the couplings
$\lambda_{SO}$ and $T_{SO}$ defined in eq.~(\ref{EQ:SOO}), the
superpotential $W_\Sigma$ and the soft SUSY-breaking terms that
involve only the adjoint fields are not relevant to the calculation of
the two-loop $\oatas$ corrections to the Higgs masses presented in
this paper, apart from contributing to the masses and mixing of the
adjoint fields as discussed in sections \ref{sec:adjoints} and
\ref{sec:gauginos} above.

In the case of the MRSSM, we must add two superfields $\mathbf{R_u}$
and $\mathbf{R_d}$ with the same gauge quantum numbers as
$\mathbf{H_d}$ and $\mathbf{H_u}$, respectively, but with different
charges under a conserved $R$-symmetry. The superpotential reads
\bea
W
&=& W_{\mathrm{Yukawa}} \,+\, W_{\mathrm{MRSSM}}~, \\[2mm]
W_{\mathrm{MRSSM}}  &=& 
(\mu_d + \lambda_{S_d}\,\mathbf{S})\,  \mathbf{R_d}\cdot  \mathbf{H_d} 
\,+\, \lambda_{T_d}\, \mathbf{H_d}\cdot  
\mathbf{T}^a \,\sigma^a\,\mathbf{R_d}\nn\\
&+& (\mu_u + \lambda_{S_u}\,\mathbf{S})\,  \mathbf{H_u}\cdot  \mathbf{R_u} 
\,+\, \lambda_{T_u}\, \mathbf{R_u}\cdot  
\mathbf{T}^a \,\sigma^a\,\mathbf{H_u}~,
\eea
while all terms involving only the MSSM-like Higgs superfields and/or
the adjoint superfields, such as those in eq.~(\ref{WDGSSM}), are
forbidden by the $R$-symmetry. The most general soft SUSY-breaking
Lagrangian for the MRSSM contains non-holomorphic mass terms for all
of the scalars, plus all of the holomorphic terms involving only the
MSSM-like Higgs scalars and/or the adjoint scalars (which, as
mentioned above, have no equivalent in the superpotential). In
contrast, the $R$-symmetry forbids Majorana mass terms for the
gauginos, and holomorphic terms for the scalars with the same
structure as the terms in the MRSSM superpotential. The requirement
that the $R$-symmetry is conserved also means that the scalar doublets
$R_u$ and $R_d$ do not develop a vacuum expectation value (vev), and
do not mix with either the MSSM-like Higgs scalars or the adjoint
scalars.

\section{Two-loop corrections in the effective potential approach}
\label{sec:2loopcalc}
\allowdisplaybreaks

In this section we adapt to the calculation of two-loop corrections to
the neutral Higgs masses in Dirac-gaugino models the
effective-potential techniques developed in
ref.~\cite{Degrassi:2001yf} for the MSSM and in
ref.~\cite{Degrassi:2009yq} for the NMSSM. We start by deriving
general results valid for all variants of Dirac-gaugino extensions of
the MSSM, then we provide explicit formulae for the MDGSSM and MRSSM
models discussed in section~\ref{sec:definition}.

\subsection{General results}
\label{sec:2loopgeneral}

The effective potential for the neutral Higgs sector can be decomposed
as $V_{\mathrm{eff}} = V_0 + \Delta V$, where $\Delta V$ incorporates
the radiative corrections. We denote collectively as $\Phi_i^0$ the
complex neutral scalars whose masses we want to calculate, and split
them into vacuum expectation values $v_i$, real scalars $S_i$
and pseudoscalars $P_i$ as
\beq
\label{phii}
\Phi_i^0 ~\equiv~ v_i + \frac{1}{\sqrt{2}} (S_i + i\, P_i)~.
\eeq
Then the mass matrices for the scalar and pseudoscalar fields can be
decomposed as
\beq
\label{massmat}
\left({\cal M}^2_S\right)^{\rm eff}_{ij}  ~=~
\left({\cal M}^2_S\right)^{\rm tree}_{ij} + 
\left(\Delta {\cal M}^2_S\right)_{ij}~,~~~~~~~~~~
\left({\cal M}^2_P\right)^{\rm eff}_{ij} ~=~
\left({\cal M}^2_P\right)^{\rm tree}_{ij} + 
\left(\Delta {\cal M}^2_P\right)_{ij}~,
\eeq
and the radiative corrections to the mass matrices are
\bea
\label{massscorr} 
\left(\Delta {\cal M}^2_S\right)_{ij} &=&
-\frac{1}{\sq2}\,\frac{\delta_{ij}}{v_i} 
\left.\frac{\partial \Delta V}{\partial S_i}
\right|_{\rm min} ~+~
\left. \frac{\partial^2 \Delta V}{\partial S_i \partial S_j}
\right|_{\rm min}~,\\
\label{masspcorr} 
\left(\Delta {\cal M}^2_P\right)_{ij} &=&
-\frac{1}{\sq2}\,\frac{\delta_{ij}}{v_i} 
\left.\frac{\partial \Delta V}{\partial S_i}
\right|_{\rm min} ~+~
\left. \frac{\partial^2 \Delta V}{\partial P_i \partial P_j}
\right|_{\rm min}~, 
\eea
where $v_i$, which we assume to be real, denote the vevs of the full
radiatively-corrected potential $V_{\mathrm{eff}}$, and the
derivatives are in turn evaluated at the minimum of the potential. The
single-derivative terms in eqs.~(\ref{massscorr}) and
(\ref{masspcorr}) arise when the minimum conditions of the potential,
\beq
\label{mincond}
\left.\frac{\partial  V_{\mathrm{eff}}}{\partial S_i}
\right|_{\rm min}=~0~,
\eeq
are used to remove the soft SUSY-breaking mass for a given field
$\Phi_i^0$ from the tree-level parts of the mass matrices. It is
understood that those terms should be omitted for fields that do not
develop a vev (such as, e.g., the fields $R_{u,d}$ in the MRSSM).

With a straightforward application of the chain rule for the
derivatives of the effective potential, the mass-matrix corrections in
eqs.~(\ref{massscorr}) and (\ref{masspcorr}) and the minimum
conditions in eq.~(\ref{mincond}) can be computed by exploiting the
Higgs-field dependence of the parameters appearing in $\Delta V$. We
restrict for simplicity our calculation to the so-called ``gaugeless
limit'', i.e.~we neglect all corrections controlled by the electroweak
gauge couplings $g$ and $g^\prime$. At the two-loop level, we focus on
the contributions to $\Delta V$ from top/stop loops that involve the
strong interactions. In Dirac-gaugino models, this results in
corrections to mass matrices and minimum conditions that are
proportional to $\alpha_s$ times various combinations of the top
Yukawa coupling $y_t$ with the superpotential couplings of the singlet
and triplet fields. It is therefore with a slight abuse of notation
that we maintain the MSSM-inspired habit of denoting collectively
those corrections as $\oatas$.

As detailed in refs.~\cite{Degrassi:2001yf, Degrassi:2009yq}, if we
neglect the electroweak contributions to the stop mass matrix the
parameters in the top/stop sector depend on the neutral Higgs fields
only through two combinations, which we denote as
$\mix \equiv |\mix|\,e^{i\varphi}$ and
$\mixt\equiv |\mixt|\,e^{i\tilde \varphi}$.  They enter the stop mass
matrix as
\beq
\label{stopmat}
\mathcal{M}^2_{\mathrm{stop}} 
~=~ \left( \begin{array}{cc} m_Q^2 + |\mix|^2 
& \mixt^* \\ \mixt & m_U^2  + |\mix|^2
\end{array} \right)~,
\eeq
where $m_Q^2$ and $m_U^2$ are the soft SUSY-breaking mass terms for
the stops.  While $\mix = y_t \,H_u^0$ both in the (N)MSSM and in
Dirac-gaugino models, the precise form of $\mixt$ depends on the model
under consideration and will be discussed later. For the time being,
we only assume that $\mixt$ is real at the minimum of the potential,
to prevent CP-violating contributions to the Higgs mass matrices.
The top/stop $\oas$ contribution to $\Delta V$ can then be expressed
in terms of five field-dependent parameters, which can be chosen as
follows. The squared top and stop masses
\beq
\label{topstop}
m_t^2 = |\mix|^2~,~~~~~~~
m^2_{\tilde{t}_{1,2}} = \frac{1}{2} \left[ ( m_Q^2 + m_U^2 + 2\,|\mix|^2\,) \pm 
 \sqrt{ (m_Q^2-m_U^2)^2 + 4 \, |\mixt|^2} \,\right]~, 
\eeq
a mixing angle $\ttbar$, with $0\leq \ttbar \leq \pi/2$, which
diagonalises the stop mass matrix after the stop fields have been
redefined to make it real and symmetric
\beq
\label{thetastop}
\sin 2 \,\ttbar = \frac{2\,|\mixt|}{\tu-\td}~,
\eeq
and a combination of the phases of $\mix$ and $\mixt$ that we can 
choose as
\beq
\label{phases} 
\cos\,(\varphi - \tilde{\varphi}) = 
\frac{ {\rm Re}(\mixt)\,{\rm Re}(\mix) + {\rm Im}(\mixt)\,{\rm Im}(\mix)}
{|\mixt| \, |\mix|} ~.
\eeq
Finally, the gluino masses $\mgi$ and the octet masses $\oi$ do not
depend on the Higgs background, since we neglect the singlet-octet
couplings $\lambda_{SO}$ and $T_{SO}$. In the following we will also
refer to $\theta_t$, with $-\pi/2 <\theta_t<\pi/2$, i.e.~the usual
field-independent mixing angle that diagonalises the stop mass matrix
at the minimum of the scalar potential.

We find general expressions for the top/stop contributions to the
minimum conditions of the effective potential and to the corrections
to the scalar and pseudoscalar mass matrices:
\bea
\label{tadgen}
\left.\frac{\partial \Delta V}{\partial S_i}\right|_{\rm min} &=&
\sdt \, \frac{\partial \mixt}{\partial S_i}\,  F  
~+~ \sqrt{2}\, y_t\, m_t \,\delta_{i2} \,G~,
\eea

\newpage

\bea
\label{MSgen}
\Big(\Delta \mathcal{M}_S^2 \Big)_{ij} &=& \bigg( 
\sdt\,\frac{\partial^2 \mixt}{\partial S_i\partial S_j}
+ \frac{ 2 }{\tu-\td} \,
\frac{\partial \mixt}{\partial S_i}\,
\frac{\partial \mixt}{\partial S_j}\,
- \frac{ \sdt }{\sqrt{2}}\,\frac{\delta_{ij}}{v_i}\,
\frac{\partial \mixt}{\partial S_j}
\bigg)\,F \nn\\
& +& \!\!\! 2 \, y_t^2\, m_t^2\, \delta_{i2} \delta_{j2}\, F_1 
+ \sqrt{2}\, m_t\, y_t\, \sdt
\bigg( \delta_{i2} \frac{\partial \mixt}{\partial S_j}\,
+ \delta_{j2} \frac{\partial \mixt}{\partial S_i} 
\bigg)\,F_2 + \sdt^2 \,
\frac{\partial \mixt}{\partial S_i}\,
\frac{\partial \mixt}{\partial S_j}\,F_3\,,\nn\\\\
\label{MPgen}
\Big(\Delta \mathcal{M}_P^2 \Big)_{ij} &=&
\bigg( \frac{ 1}{\tu-\td} 
\frac{\partial^2 |\mixt|^2}{\partial P_i \partial P_j} 
- \frac{\sdt}{\sqrt{2}} \, \frac{\delta_{ij}}{v_i} \, 
\frac{\partial \mixt}{\partial S_j} \bigg)\,F\nn \\
&+& \bigg( \frac{\delta_{i2}}{v_2}\,  \mixt
+ \sqrt{2} \, i \,\frac{\partial \mixt}{\partial P_i} \bigg) 
\bigg( \frac{ \delta_{j2}}{v_2} \,  \mixt
+ \sqrt{2} \, i \,\frac{\partial \mixt}{\partial P_j}\bigg) \, 
\tan\beta\,F_\varphi~,
\eea
where all quantities are understood as evaluated at the minimum of the
potential, no summation is implied over repeated indices, the fields
are ordered as
$(\Phi_1^0\,,\Phi_2^0\,,\,...) = (H_d^0\,,H_u^0\,,\,...)\,$, and again
the terms involving $\delta_{ij}/v_i$ should be omitted if $\Phi_i^0$
does not develop a vev. The angle $\beta$ is defined as in the MSSM by
$\tan\beta = v_2/v_1$.  Here and thereafter we also adopt the
shortcuts $c_\phi \equiv \cos\phi$ and $s_\phi \equiv \sin\phi$ for a
generic angle $\phi$.  The functions $F, G, F_1, F_2, F_3$ and
$F_\varphi$ entering eqs.~(\ref{tadgen})--(\ref{MPgen}) are
combinations of the derivatives of $\Delta V$. Explicit expressions
for most of those functions can be found e.g.~in
ref.~\cite{Degrassi:2009yq}, but we display all of them here for
completeness:
\bea
\label{defF}
F & = & \DVtu - \DVtd - \frac{4\,\cdt^2}{\tu-\td}\,\DVcdtq~,\\[2.5mm]
\label{defG}
G & = & \DVt + \DVtu + \DVtd~,\\[2.5mm]
\label{defF1}
F_1 & = &
\DVtt + \DVtutu + \DVtdtd + 2\,\DVttu + 2\,\DVttd + 2\,\DVtutd ~,\\[2.5mm]
F_2 & = &
\DVtutu - \DVtdtd + \DVttu - \DVttd \nn\\
\label{defF2}
&& - \frac{4 \,\cdt^2}{\tu-\td}
\left( \DVcdtqt+\DVcdtqtu+\DVcdtqtd\right)~,\\[2.5mm]
\label{defF3}
F_3 & = &
\DVtutu + \DVtdtd -2\,\DVtutd - \frac{2}{\tu-\td}\left(\DVtu-\DVtd\right)\nn\\
&& + \frac{ 16\, \cdt^2}{(\tu-\td)^2}\,\left( \cdt^2\,\DVcdtqcdtq 
+2\,\DVcdtq \right)
- \frac{ 8\, \cdt^2}{\tu-\td}\,\left( \DVcdtqtu-\DVcdtqtd \right)~,\nn\\
\\
F_\varphi &=& -\frac{2\,z_t\,\cot\beta}{\sdt^2\,(\tu-\td)^2}\,
\frac{\partial \Delta V}{\partial c_{\varphi_t - \tilde{\varphi}_t}}~,
\eea
where we defined $z_t \equiv {\rm sign}(\mixt|_{\rm min})$.  

\subsection{Two-loop top/stop contributions to the effective potential}

\begin{figure}[t]
\begin{center}
\includegraphics[width=3cm]{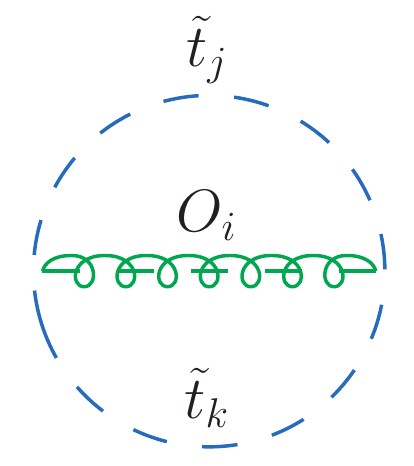}
\end{center}
\vspace*{-5mm}
\caption{Novel two-loop contribution to the effective potential
  involving stops and octet scalars.}
\label{fig:diag}
\end{figure}

For the computation of the two-loop $\oatas$ corrections to the Higgs
mass matrices in models with Dirac gauginos we need the explicit
expression for the top/stop $\oas$ contribution to $\Delta V$,
expressed in terms of the field-dependent parameters defined in the
previous section. In addition to the contributions of diagrams
involving gluons, gluinos or the D-term-induced quartic stop
couplings, which are in common with the (N)MSSM and can be found in
ref.~\cite{Degrassi:2001yf}, $\Delta V$ receives a contribution from
the diagram shown in figure~\ref{fig:diag}, involving stops and octet
scalars.

We assume that the gaugino masses are real so that the diagonalising
matrix $R_{ij}$ is real and $R_{1i}^2$ is positive, but allow $\mgi$
to be negative. Since $R_{11}^2 + R_{12}^2 = 1$, we can simply write
the top/stop $\oas$ contribution to the two-loop effective potential
(in units of $\as\, C_F N_c\,/(4 \pi)^3$, where $C_F = 4/3$ and $N_c =
3$ are colour factors) as
\beq
\label{V2loop}
\Delta V^{\,\as} 
~=~  \sum_{i=1}^2 \,R_{1i}^2 \,\Delta V^{\,\as}_{\smallMSSM}
\,
+~~ \Delta V^{\,\as}_{\rm octet}~, 
\eeq
where $\Delta V^{\,\as}_{\smallMSSM}$ is the analogous contribution in the
(N)MSSM,
\bea
\label{v2as}
\Delta V^{\,\as}_{\smallMSSM} & = & 
2\, J(\t,\t) - 4\,\t\,I(\t,\t,0) + \nn \\
&+& 
\biggr\{ 2\,\tu\,I(\tu,\tu,0) + 2\,L(\tu,\gi,\t)
- 4\,m_t\,\mgi\,s_{2\bar\theta}\,c_{\varphi - \tilde{\varphi}}
\,I(\tu,\gi,\t) \nn\\
&& +\,\frac12\,  (1+c_{2\bar\theta}^2)\,J(\tu,\tu) 
+ \frac{s_{2\bar\theta}^2}{2} J(\tu,\td)\;\; 
+ \;\; \left[ \tul \leftrightarrow \tdl\,,\,
s_{2\bar\theta} \rightarrow - s_{2\bar\theta}\right] \biggr\},~~~
\eea
while $\Delta V^{\,\as}_{\rm octet}$ is the additional $\oas$
contribution of the two-loop diagram shown in figure~\ref{fig:diag},
involving stops and octet scalars. The latter can be decomposed as
\beq
\label{Voctet}
\Delta V^{\,\as}_{\rm octet} ~\equiv~ m_D^2\,\left( 
c_{\phi_O}^2 \Delta V_{O_1} ~+~ s_{\phi_O}^2 \Delta V_{O_2}
\right)~,
\eeq
with
\beq
\label{Voctet_Oi}
\Delta V_{O_i} ~=~ -2 \,\cdtt^2 \,\left[ I(\tu,\tu,\oi) 
\,+\,I(\td,\td,\oi)\right] 
\,-\, 4\,\sdtt^2\,I(\tu,\td,\oi)~.
\eeq
The two-loop integrals $J(x,y)$, $I(x,y,z)$ and $L(x,y,z)$ entering
eqs.~(\ref{v2as}) and (\ref{Voctet_Oi}) are defined, e.g., in
eqs.~(D1)--(D3) of ref.~\cite{Degrassi:2009yq}, and were first
introduced in ref.~\cite{Ford:1992pn}.
Explicit expressions for the derivatives of $\Delta V^{\,\as}$, valid
for all Dirac-gaugino models considered in this paper, are provided in
appendix~\ref{sec:2lderivs}.

We remark that, by using the ``minimally subtracted'' two-loop
integrals of ref.~\cite{Ford:1992pn}, we are implicitly assuming a
$\drbar$ renormalisation for the parameters entering the tree-level
and one-loop parts of the effective potential. Consequently, our
results for the two-loop top/stop contributions to mass matrices and
minimum conditions also assume that the corresponding tree-level and
one-loop parts are expressed in terms of $\drbar$-renormalised
parameters. We will describe in section~\ref{sec:OS} how our two-loop
formulae should be modified if the top/stop parameters entering the
one-loop part of the corrections are expressed in a different
renormalisation scheme. For what concerns the parameters entering the
tree-level mass matrices for scalars and pseudoscalars -- whose
specific form depends on the Dirac-gaugino model under consideration
-- they can be taken directly as $\drbar$-renormalised inputs at some
reference scale $Q$, at least in the absence of any experimental
information on an extended Higgs sector. Exceptions are given by the
electroweak gauge couplings and by the combination of doublet vevs
$v \equiv (v_1^2+ v_2^2)^{1/2}\,$, which in general should be
extracted from experimentally known observables such as, e.g., the
muon decay constant and the gauge-boson masses. As was pointed out for
the NMSSM in ref.~\cite{Muhlleitner:2014vsa}, the extraction of the
$\drbar$ parameter $v(Q)$ involves two-loop corrections whose effects
on the scalar and pseudoscalar mass matrices are formally of the same
order as some of the $\oatas$ corrections computed in this
paper\footnote{These additional $\oatas$ effects arise from terms in
  the tree-level mass matrices in which $v$ appears in combination
  with the singlet or triplet superpotential couplings. In contrast,
  in the MSSM all occurrences of $v$ in the tree-level mass matrices
  are multiplied by the electroweak gauge couplings, thus they are not
  relevant in the gaugeless limit.}. However, a two-loop determination
of $v(Q)$ goes beyond the scope of our calculation, as it requires
two-loop contributions to the gauge-boson self-energies which cannot
be obtained with effective-potential methods. Besides,
ref.~\cite{Muhlleitner:2014vsa} showed that, at least in the NMSSM
scenarios considered in that paper, the $\oatas$ effects on the scalar
masses arising from the two-loop corrections to $v$ are quite small,
typically of the order of a hundred MeV.

\subsection{Mass corrections in the MDGSSM}
\label{sec:MDGSSM}

The MDGSSM contains a singlet $S$ and an $SU(2)$ triplet $T^a$ which
mix with the usual Higgs fields $H_d$ and $H_u$. In this model, the
stop mixing term $\mixt$ defined in eq.~(\ref{stopmat}) reads
\beq
\label{stopmix}
\mixt ~=~ y_t\,\biggr(A_t\,H_u^0 - \mu \,H_d^{0\,*} - \lambda_S \,S^*\, H_d^{0\,*}
- \lambda_T \,T^{0\,*}\,H_d^{0\,*}\biggr)~,
\eeq
where $A_t$ is the soft SUSY-breaking trilinear interaction term for Higgs
and stops.  We order the neutral components of the fields as $\Phi_i^0
= (H_d^0,\, H_u^0,\,S,\,T^0)$ and expand them as in
eq.~(\ref{phii}). For the minimum conditions of the effective
potential, eq.~(\ref{tadgen}) gives
\bea
\label{tad_MDGSSM}
\left.\frac{\partial \Delta V}{\partial S_1}\right|_{\rm min}
&=& - y_t \,\frac{\tilde{\mu}}{\sqrt{2}} \,\sdt\, F~,\\[1mm]
\left.\frac{\partial \Delta V}{\partial S_2}\right|_{\rm min}
&=& \sq2\,y_t\,\mt\, G ~+~ y_t \,\frac{A_t}{\sqrt{2}}\,\sdt\, F~,\\[1mm]
\left.\frac{\partial \Delta V}{\partial S_3}\right|_{\rm min}
&=& - y_t \,\frac{\lambda_S\,v_1}{\sqrt{2}} \,\sdt\, F~,\\[1mm]
\left.\frac{\partial \Delta V}{\partial S_4}\right|_{\rm min}
&=& - y_t \,\frac{\lambda_T\,v_1}{\sqrt{2}} \,\sdt\, F~,
\eea
where we defined $\tilde\mu \,\equiv\, \mu + \lambda_S\,v_3 + \lambda_T\,v_4$.
For the corrections to the mass matrices of scalars and pseudoscalars,
eqs.~(\ref{MSgen}) and (\ref{MPgen}) give
\bea
\Big(\Delta\mathcal{M}^2_S\Big)_{11} &=& 
\frac{1}{2}\, y_{t }^{2}\,\tilde{\mu}^{2}\, s_{2\theta_t}^{2}  \,F_3
~+ ~\frac{y_{t }^{2}\,A_t \, \tilde{\mu} \,\tan \beta  }{\tu-\td}\,F~,\\[1mm]
\Big(\Delta\mathcal{M}^2_S\Big)_{12} &=& 
- \,y_{t }^{2}\, m_t \, \tilde{\mu} \, s_{2\theta_t}  \, \,F_2
~-~\frac{1}{2} \,y_{t }^{2}\,A_t \, \tilde{\mu} \, s_{2\theta_t}^{2}  \,F_3
~-\, \frac{ y_{t }^{2} \, A_t \, \tilde{\mu} }{\tu-\td}\,F~,\\[1mm]
\Big(\Delta\mathcal{M}^2_S\Big)_{22} &=& 
2 \, y_{t }^{2}\,m_{t }^{2}  \,F_1
~+~ 2\, y_{t }^{2}\,  m_t \,A_t \, s_{2\theta_t}  \,F_2
~+~ \frac{1}{2}\, y_{t }^{2}\, A_{t }^{2} \,s_{2\theta_t}^{2}  \,F_3
~+ \,\frac{ y_{t }^{2} A_t \, \tilde{\mu} \, \cot \beta  }{\tu-\td}\,F~,\\[1mm]
\Big(\Delta\mathcal{M}^2_S\Big)_{13} &=&
\frac{1}{2}\, y_t \, \lambda_S \, m_t \, \tilde{\mu} \,\cot\beta\,
s_{2\theta_t}^{2} \,F_3
~-\,\frac{ y_t\,  \lambda_S\,  m_t \,\Big(A_t-2\, \tilde{\mu}\,\cot\beta \Big)}
{\tu-\td}\,F~,\\[1mm]
\Big(\Delta\mathcal{M}^2_S\Big)_{23} &=& 
\!\!- y_t \, \lambda_S\, m_{t }^{2}\,  \cot \beta\, s_{2\theta_t}   \,F_2
-\frac{1}{2}\, y_t \, \lambda_S \, A_t \, m_t \cot \beta 
\, s_{2\theta_t}^{2}  \,F_3
- \frac{ y_t  \, \lambda_S \,m_t \,A_t  \, \cot \beta}{\tu-\td}\,F ,\\[1mm]
\Big(\Delta\mathcal{M}^2_S\Big)_{33} &=&
\frac{1}{2}\,\lambda_{S }^{2}\,  
m_{t }^{2} \,  {\cot}^{2}\beta\, s_{2\theta_t}^{2}\,F_3
~+ ~\,\frac{ \lambda_S \,m_{t }^{2} \, \cot \beta\, 
\Big(A_t +( \lambda_S\,v_3- \tilde{\mu})\cot\beta\Big)}
{v_3\,(\tu-\td)}\,F~,\\[1mm]
\Big(\Delta\mathcal{M}^2_S\Big)_{14} &=& 
\frac{1}{2}\, y_t \,
\lambda_T\,  m_t\, \tilde{\mu}\,  \cot \beta\, s_{2\theta_t}^{2}    \,F_3
~-\,\frac{  y_t\,  \lambda_T\, m_t \,\Big(A_t-2\, \tilde{\mu} \, \cot\beta\Big)}
{\tu-\td}\,F~,\\[1mm]
\Big(\Delta\mathcal{M}^2_S\Big)_{24} &=& 
\!\!- y_t \, \lambda_T\, m_{t }^{2}\,  \cot \beta\, s_{2\theta_t}   \,F_2
-\frac{1}{2}\, y_t \, \lambda_T \, A_t \, m_t \cot \beta 
\, s_{2\theta_t}^{2}  \,F_3
- \frac{ y_t  \, \lambda_T \,m_t \,A_t  \, \cot \beta}{\tu-\td}\,F ,\\[1mm]
\Big(\Delta\mathcal{M}^2_S\Big)_{34} &=& 
\frac{1}{2}\, \lambda_S \, \lambda_T\, m_{t }^{2}\,{\cot}^{2}\beta\,\sdt^2\,F_3
~+\,\frac{ \lambda_S\, \lambda_T\, m_{t }^{2}\, 
{\cot}^{2} \beta }{\tu-\td}\,F~,\\[1mm]
\Big(\Delta\mathcal{M}^2_S\Big)_{44} &=&
\frac{1}{2}\,\lambda_{T}^{2}\,  
m_{t }^{2} \,  {\cot}^{2}\beta\, s_{2\theta_t}^{2}\,F_3
~+ ~\,\frac{ \lambda_T \,m_{t }^{2} \, \cot \beta\, 
\Big(A_t +( \lambda_T\,v_4- \tilde{\mu})\cot\beta\Big)}
{v_4\,(\tu-\td)}\,F~,\\[5mm]
\Big(\Delta\mathcal{M}^2_P\Big)_{11} &=&
\frac{y_{t }^{2} \, A_t\,\tilde{\mu}\,  \tan \beta}{\tu-\td}\,F~+~
y_{t }^{2}\, \tilde{\mu}^{2} \,\tan \beta  \,F_{\varphi}~,\\[1mm]
\Big(\Delta\mathcal{M}^2_P\Big)_{12} &=&
\frac{y_{t }^{2} \, A_t\,\tilde{\mu}}{\tu-\td}\,F~+~
y_{t }^{2}\, \tilde{\mu}^{2}  \,F_{\varphi}~,\\[1mm]
\Big(\Delta\mathcal{M}^2_P\Big)_{22} &=&
\frac{y_{t }^{2} \, A_t\,\tilde{\mu}\,  \cot \beta}{\tu-\td}\,F~+~
y_{t }^{2}\, \tilde{\mu}^{2} \,\cot \beta  \,F_{\varphi}~,\\[1mm]
\Big(\Delta\mathcal{M}^2_P\Big)_{13} &=&
\frac{y_t\,\lambda_S \, \mt\,A_t}{\tu-\td}\,F~+~
y_t\,\lambda_S \, \mt\,\tilde{\mu}  \,F_{\varphi}~,\\[1mm]
\Big(\Delta\mathcal{M}^2_P\Big)_{23} &=&
\frac{y_t\,\lambda_S \, \mt\,A_t\,  \cot \beta}{\tu-\td}\,F~+~
y_t\,\lambda_S \, \mt\,\tilde{\mu} \,  \cot \beta \,F_{\varphi}~,\\[1mm]
\Big(\Delta\mathcal{M}^2_P\Big)_{33} &=&
\frac{ \lambda_S \,m_{t }^{2} \, \cot \beta\, 
\Big(A_t +( \lambda_S\,v_3- \tilde{\mu})\cot\beta\Big)}
{v_3\,(\tu-\td)}\,F ~+~
\lambda_S^2\,\mt^2\,\cot\beta\,F_{\varphi}~,\\[1mm]
\Big(\Delta\mathcal{M}^2_P\Big)_{14} &=&
\frac{y_t\,\lambda_T \, \mt\,A_t}{\tu-\td}\,F~+~
y_t\,\lambda_T \, \mt\,\tilde{\mu}  \,F_{\varphi}~,\\[1mm]
\Big(\Delta\mathcal{M}^2_P\Big)_{24} &=&
\frac{y_t\,\lambda_T \, \mt\,A_t\,  \cot \beta}{\tu-\td}\,F~+~
y_t\,\lambda_T \, \mt\,\tilde{\mu} \,  \cot \beta \,F_{\varphi}~,\\[1mm]
\Big(\Delta\mathcal{M}^2_P\Big)_{34} &=&
\frac{\lambda_S\,\lambda_T \, \mt^2\, \cot^2\beta}{\tu-\td}\,F~+~
\lambda_S\,\lambda_T \, \mt^2 \,  \cot \beta \,F_{\varphi}~,\\[1mm]
\Big(\Delta\mathcal{M}^2_P\Big)_{44} &=&
\frac{ \lambda_T \,m_{t }^{2} \, \cot \beta\, 
\Big(A_t +( \lambda_T\,v_4- \tilde{\mu})\cot\beta\Big)}
{v_4\,(\tu-\td)}\,F ~+~
\lambda_T^2\,\mt^2\,\cot\beta\,F_{\varphi}~.
\eea

\subsection{Mass corrections in the MRSSM}
\label{sec:MRSSM}

The MRSSM is defined to be $R$-symmetric, and has fields $R_u, R_d$
which pair with the Higgs fields without themselves developing
vevs. In this model the gluino mass terms are purely Dirac, therefore,
in our conventions, $R_{11}^2=R_{12}^2=1/2\,$ and $m_{\tilde g_1} =
-m_{\tilde g_2} = m_D\,$.  The trilinear Higgs-stop coupling $A_t$ is
forbidden, and the term $\mixt$ defined in eq.~(\ref{stopmat}) reads
\beq
\mixt ~=~ -y_t\,\biggr(\mu_u + \lambda_{S_u} \,S^*
+ \lambda_{T_u} \,T^{0\,*}\biggr)\,R_u^{0\,*}~,
\eeq
and vanishes at the minimum of the scalar potential, hence the stops
do not mix.  Moreover, the term proportional to $c_{\varphi -
  \tilde{\varphi}}$ in the second line of eq.~(\ref{v2as}) cancels out
in the sum over the gluino masses.  As a consequence, the radiative
corrections induced by top/stop loops are remarkably simple. Ordering
the neutral components of the fields as $\Phi_i^0 = (H_d^0,\,
H_u^0,\,S,\,T^0,R_d^0,\, R_u^0)$, we find that the only non-vanishing
contributions to the minimum conditions of the potential and to the
Higgs mass matrices are
\bea
\left.\frac{\partial \Delta V}{\partial S_2}\right|_{\rm min}
&=& \sq2\,y_t\,\mt\, G~,\\[2mm]
\Big(\Delta\mathcal{M}^2_S\Big)_{22} &=& 
2 \, y_{t }^{2}\,m_{t }^{2}  \,F_1~,\\[2mm]
\Big(\Delta\mathcal{M}^2_S\Big)_{66} &=& 
\Big(\Delta\mathcal{M}^2_P\Big)_{66} ~=~ 
\frac{y_t^2\,\tilde \mu_u^2}{\tu-\td}\,F~,
\eea
where we defined $\tilde \mu_u\,\equiv\, \mu_u + \lambda_{S_u}\,v_3 +
\lambda_{T_u}\,v_4$.

\subsection{On-shell parameters in the top/stop sector}
\label{sec:OS}

The results presented so far for the two-loop corrections to the
neutral Higgs masses in models with Dirac gauginos were obtained under
the assumption that the parameters entering the tree-level and
one-loop parts of the mass matrices are renormalised in the $\drbar$
scheme. While this choice allows for a straightforward implementation
of our results in automated calculations such as the one of \SARAH, it
is well known that, in the $\drbar$ scheme, the Higgs-mass calculation
can be plagued by unphysically large contributions if there is a
hierarchy between the masses of the particles running in the
loops~\cite{Degrassi:2001yf}. In particular, the contributions of
two-loop diagrams involving stops and gluinos include terms
proportional to $m^2_{\tilde g_i}/m^2_{\tilde t_j}$, which can become
very large in scenarios with gluinos much heavier than the
stops. Since this kind of hierarchy can occur naturally (i.e., without
excessive fine tuning in the squark masses) in scenarios with Dirac
gluino masses~\cite{Fox:2002bu}, it is useful to re-express the
one-loop part of the corrections to the Higgs masses in terms of
OS-renormalised top/stop parameters. In that case, the terms
proportional to $m^2_{\tilde g_i}$ in the two-loop part of the
corrections cancel out against analogous contributions induced by the
OS counterterms, leaving only a milder logarithmic dependence of the
Higgs masses on the gluino masses.

Since we are focusing on the $\oatas$ corrections to the Higgs masses,
we need to provide an OS prescription only for parameters in the
top/stop sector that are subject to $\oas$ corrections, i.e.~$\mt$,
$\tu$, $\td$ and $\theta_t$.  In models that allow for a trilinear
Higgs-stop coupling $A_t$ -- such as the MDGSSM, see
eq.~(\ref{stopmix}) -- its counterterm can be derived from those of
the other four parameters via the relation
$(\tu-\td)\,\sin 2\theta_t = 2 \,\widetilde X|_{\rm min}\,$ (in
general, the stop mixing $\widetilde X|_{\rm min}$ contains other
terms in addition to $\mt\,A_t$, but they are exempt from $\oas$
corrections). Finally, since the vevs $v_i$ are not renormalised at
$\oas$, the top Yukawa coupling $y_t$ receives the same relative
correction as the top mass. Defining
$x_k^{\smalldrbar} = x_k^{\smallOS} + \delta x_k$ for each parameter
$x_k \equiv (\mt,\,\tu,\,\td,\,\theta_t,\,A_t)$, the $\drbar\,$--\,OS
shifts of top and stop masses and mixing are given in terms of the
finite parts (here denoted by a hat) of the top and stop self-energies
\beq
\label{counter}
\delta m_t \,=\, \hat \Sigma_t(\mt)~,~~~~~~
\delta \ti \,=\, \hat \Pi_{ii}(\ti)~~~(i=1,2),~~~~~~
\delta \theta_t \,=\, \frac12 \,\frac{\hat\Pi_{12}(\tu)+\hat \Pi_{12}(\td)}
{\tu-\td}~,
\eeq
and the shift for the trilinear coupling reads
\beq
\label{dAt}
\delta A_t ~=~ 
\left( \frac{\delta \tu - \delta\td}{\tu-\td}\,-\, \frac{\delta\mt}{\mt}
+ 2\,\cot2\theta_t\,\delta\theta_t
\right)\widetilde X|_{\rm min}~~.
\eeq
As in the case of the two-loop effective potential in eq.~(\ref{V2loop}), the 
$\drbar\,$--\,OS shifts $\delta x_k$ can be cast as
\beq
 \delta x_k
~=~  \sum_{i=1}^2 \,R_{1i}^2 \,(\delta x_k^{\smallMSSM})_i
\,+~~ \delta x_k^{\rm octet}~, 
\eeq
where $(\delta x_k^{\smallMSSM})_i$ are obtained, with the trivial
replacement $\mg \rightarrow \mgi$, from the MSSM shifts given in
appendix B of ref.~\cite{Degrassi:2001yf}, whereas $\delta x_k^{\rm
  octet}$ are novel contributions involving the octet scalar. In
particular, $\delta \mt^{\rm octet}=0$, and the remaining shifts can
be obtained by combining as in eqs.~(\ref{counter}) and (\ref{dAt})
the octet contributions to the finite parts of the stop self-energies:
\bea
\hat\Pi_{11}(\tu)^{\rm octet}\!\!&=&
\frac{g_s^2\,m_D^2}{4 \pi^2}\,C_F\,c_{\phi_O}^2
\left[\cdt^2\,\hat B_0(\tu,\tu,m_{O_1}^2) 
+ \sdt^2 \,\hat B_0(\tu,\td,m_{O_1}^2)\right]\nonumber\\[2mm]
&+&~(c_{\phi_O}\rightarrow s_{\phi_O}\,,~ 
m_{O_1}\rightarrow m_{O_2})\,,\\[3mm]
\hat\Pi_{22}(\td)^{\rm octet}\!\!&=&
\frac{g_s^2\,m_D^2}{4 \pi^2}\,C_F\,c_{\phi_O}^2
\left[\cdt^2\,\hat B_0(\td,\td,m_{O_1}^2) 
+ \sdt^2 \,\hat B_0(\td,\tu,m_{O_1}^2)\right]\nonumber\\[2mm]
&+&~(c_{\phi_O}\rightarrow s_{\phi_O}\,,~ 
m_{O_1}\rightarrow m_{O_2})\,,\\[3mm]
\hat\Pi_{12}(p^2)^{\rm octet}\!\!&=&
-\frac{g_s^2\,m_D^2}{4 \pi^2}\,C_F\,c_{\phi_O}^2\,\cdt\,\sdt\,
\left[\hat B_0(p^2,\tu,m_{O_1}^2) - \hat B_0(p^2,\td,m_{O_1}^2)\right]
\nonumber\\[2mm]
&+&~(c_{\phi_O}\rightarrow s_{\phi_O}\,,~ 
m_{O_1}\rightarrow m_{O_2})~,
\eea
where $\hat B_0(p^2,m_1^2,m_2^2)$ is the finite part of the
Passarino-Veltman function.

The change in renormalisation scheme for the top/stop parameters
entering the one-loop ($1\ell$) part of the corrections to the Higgs
mass matrices induces a shift in the two-loop ($2\ell$) part of the
corrections:
\beq
\label{2lshifts}
\delta\left(\Delta {\cal M}^2_{S,P}\right)^{2\ell}_{ij} ~=~
\sum_k~ \delta x_k\,\frac{\partial}{\partial x_k} 
\left(\Delta {\cal M}^2_{S,P}\right)^{1\ell}_{ij}~~.
\eeq
Analogous expressions hold for the shifts in the two-loop part of the
minimum conditions of the effective potential. The one-loop
corrections entering the equation above can be obtained by inserting
in eqs.~(\ref{tadgen})--(\ref{MPgen}) the one-loop expressions for the
functions $F$, $G$, $F_{1,2,3}$ and $F_\varphi\,$. In units of
$N_c/(16\pi^2)$, these read:
\beq
\nonumber
F^{1\ell} ~=~ \tu\left(\ln\frac\tu {Q^2}-1\right)\,-~
\td\left(\ln\frac\td {Q^2}-1\right)~,
\eeq
\beq
\nonumber
G^{1\ell} ~=~ \tu\left(\ln\frac\tu {Q^2}-1\right)\,+~
\td\left(\ln\frac\td {Q^2}-1\right)
\,-~2\,\t\left(\ln\frac\t {Q^2}-1\right)~,
\eeq
\beq
F_1^{1\ell}~=~\ln\frac{\tu\td}{\mt^4}~,~~~~
F_2^{1\ell}~=~\ln\frac{\tu}{\td}~,~~~~
F_3^{1\ell}~=~\left(2-\frac{\tu+\td}{\tu-\td}\,\ln\frac{\tu}{\td}\right)\,,~~~~
F_\varphi^{1\ell}~=~0~,
\label{eq:F1loop}
\eeq

\noindent where $Q$ is the renormalisation scale at which the
parameters entering the tree-level and one-loop parts of the mass
matrices are expressed. As mentioned above, the $\drbar\,$--\,OS
shifts derived in eq.~(\ref{2lshifts}) cancel the power-like
dependence of the two-loop corrections on the gluino masses.

\subsection{Obtaining the $\oabas$ corrections}

Our $\drbar$ computation of the $\oatas$ corrections allows us to
obtain also the two-loop $\oabas$ corrections induced by the
bottom/sbottom sector, which can be relevant for large values of
$\tan\beta$. To this purpose, the substitutions $t\rightarrow b$,
$u\rightarrow d$,
$\partial \Delta V/\partial S_1 \leftrightarrow \partial \Delta
V/\partial S_2$,
$\left(\Delta {\cal M}^2_{S,P}\right)_{11} \leftrightarrow ~
\left(\Delta {\cal M}^2_{S,P}\right)_{22}$,
$\left(\Delta {\cal M}^2_{S,P}\right)_{1k} \leftrightarrow ~
\left(\Delta {\cal M}^2_{S,P}\right)_{2k}$ (with $k>2$) and
$\tan\beta \leftrightarrow \cot\beta$ must be performed in the
formulae of sections~\ref{sec:MDGSSM} and~\ref{sec:MRSSM}. In the case
of the bottom/sbottom corrections, however, passing from the $\drbar$
scheme to the OS scheme would involve additional complications, as
explained in ref.~\cite{Brignole:2002bz}.

\subsection{Simplified formulae}
\label{sec:simpli}

Having computed the general expressions for the two-loop corrections
to the neutral Higgs masses in models with Dirac gauginos, it is now
interesting to provide some approximate results for the dominant
corrections to the mass of a SM-like Higgs. We focus on the case of
a purely-Dirac mass term for the gluinos, which -- as mentioned
earlier -- implies that we can set $R_{11}^2=R_{12}^2=1/2\,$ and
$m_{\tilde g_1} = -m_{\tilde g_2} = \mg\,$, with $\mg \equiv m_D$.
We also restrict ourselves to the decoupling limit in which all
neutral states except a combination of $H_d^0$ and $H_u^0$ are heavy,
so that
\beq
H_d^0 ~\approx\, \left( v + \frac{h}{\sqrt{2}} \right) \,\cos \beta ~+~ ...\,,
~~~~~~~~
H_u^0 ~\approx\, \left( v + \frac{h}{\sqrt{2}} \right)  \,\sin \beta ~+~ ...\,, 
\eeq
where $v\approx 174$~GeV, and all other fields have negligible mixing
with the lightest scalar $h$, which is SM-like.    We can
then approximate the correction to the squared mass $m_h^2$ as
\beq
\Delta m_h^2 ~\approx~ \cos^2 \beta\, \left(\Delta \mathcal{M}_S^2\right)_{11} 
~+~ \sin^2 \beta \,\left( \Delta \mathcal{M}_S^2\right)_{22} 
~+~ \sin 2\beta  \,\left(\Delta \mathcal{M}_S^2\right)_{12}.
\eeq
Finally, we assume
that the superpotential couplings of the adjoint fields (e.g., the
couplings $\lambda_S$ and $\lambda_T$ in the MDGSSM) are subdominant
with respect to the top Yukawa coupling, so that we can focus on the
two-loop corrections proportional to $\alpha_s\,m_t^4/v^2$.

With these restrictions, we shall give useful formulae valid for a
phenomenologically interesting subspace of \emph{all} extant Dirac
gaugino models; while in the following we refer to simplified MDGSSM
and MRSSM scenarios, this merely reflects whether stop mixing is
allowed.

\subsubsection{Common SUSY-breaking scale}
\label{sec:commonMS}
We first consider a simplified MDGSSM scenario in which the soft
SUSY-breaking masses for the two stops and the Dirac mass of the
gluinos are large and degenerate, i.e.~$m_Q = m_U = \mg = \msusy$ with
$\msusy \gg \mt$. Expanding our result\,\footnote{We have verified
  that, for $\msusy = 1$~TeV and for $|\hat X_t|$ up to the ``maximal
  mixing'' value of $\sqrt 6$, the predictions for $m_h$ obtained with
  the simplified formulae of this section agree at the per-mil level
  with the unexpanded result. For larger $\msusy$ the accuracy of our
  approximation improves, and for $|\hat X_t|>\sqrt 6$ it degrades.}
for the top/stop contributions to $\Delta m_h^2$ at the leading order
in $\mt/\msusy$, we can decompose it as
\beq
\label{dmh_MS}
\Delta m_h^2 ~\approx~ \frac{3\,m^4_t}{4\,\pi^2 v^2}\,
\left[\,\ln\frac{\msusy^2}{m_t^2} + {\hat X_t^2}
-\frac{\hat X_t^4}{12}\,\right]
~+~\left(\Delta m_h^2\right)_{2\ell}^{{\scriptscriptstyle {\rm ``MSSM"}}} 
+~ c^2_{\phi_{O}} \left(\Delta m_h^2\right)_{2\ell}^{O_1}
~+~ s^2_{\phi_{O}} \left(\Delta m_h^2\right)_{2\ell}^{O_2}~,
\eeq
where $\hat X_t \equiv X_t/\msusy$, in which
$X_t = A_t - \tilde \mu\cot\beta\,$ is the left-right mixing term in
the stop mass matrix with $\tilde \mu$ defined as in
section~\ref{sec:MDGSSM}. The first term in $\Delta m_h^2$ is the
dominant 1-loop contribution from diagrams with top quarks or stop
squarks, which is the same as in the MSSM.
The second term is the $\oatas$ contribution from two-loop, MSSM-like
diagrams involving gluons, gluinos or a four-stop interaction. Under
the assumption that the parameters $\mt$, $\msusy$ and $A_t$ entering
the one-loop part of the correction are renormalised in the $\drbar$
scheme at the scale $Q$, it reads
\beq
\label{eq:mssmcorrDR}
\left(\Delta m_h^2\right)_{2\ell}^{{\scriptscriptstyle {\rm ``MSSM"}}}=~
\frac{\alpha_s \mt^4}{2\,\pi^3v^2}
\left\{\ln^2\frac{\msusy^2}{m^2_t}-2\ln^2\frac{\msusy^2}{Q^2}
+2\ln^2 \frac{m^2_t}{Q^2}
+\ln\frac{\msusy^2}{m^2_t} -1 +{\hat X_t^2}\,
\left[1-2\ln\frac{\msusy^2}{Q^2}\right]
-\frac{\hat X_t^4}{12}\right\}.
\eeq
We remark that this correction differs from the usual one in the MSSM,
see e.g.~eq.~(21) of ref.~\cite{Espinosa:1999zm}, due to the absence
of terms involving odd powers of ${\hat X_t}$. Indeed, those terms are
actually proportional to the gluino masses, and in the considered
scenario they cancel out of the sum over the gluino mass eigenstates,
because $m_{\tilde g_1} = -m_{\tilde g_2}$. If the parameters $\mt$,
$\msusy$ and $A_t$ are renormalised in the OS scheme as described in
section \ref{sec:OS}, the correction reads instead
\beq
\label{eq:mssmcorrOS}
\left(\Delta m_h^2\right)_{2\ell}^{{\scriptscriptstyle {\rm ``MSSM"}}}=~
-\frac{3\,\alpha_s \mt^4}{2\,\pi^3v^2}
\left\{\ln^2\frac{\msusy^2}{m^2_t}
+\left[2+{\hat X_t^2}\right]\,\ln\frac{\msusy^2}{m^2_t}
+\frac{\hat X_t^4}{4}\right\}.
\eeq
Note that the explicit dependence on the renormalisation scale $Q$
drops out. Again, this correction differs from the usual one in the
MSSM, see e.g.~the first line in eq.~(20) of
ref.~\cite{Espinosa:2000df}, due to the absence of a term linear in
$\hat X_t$.

Finally, the last two terms on the right-hand side of
eq.~(\ref{dmh_MS}) represent the $\oatas$ contributions of two-loop
diagrams with stops and octet scalars, which are specific to models
with Dirac gluinos. In the $\drbar$ scheme they read
\bea
\left(\Delta m_h^2\right)_{2\ell}^{O_i} &=&
-\frac{\alpha_s \mt^4}{\pi^3v^2}\,\left\{ 
1 - \ln \frac{\msusy^2}{Q^2} + f(x_i)
- \hat X_t^2\, \left[1 - \ln \frac{m_{O_i}^2}{Q^2} 
+ 2\,x_i\,f(x_i)\right]\right.\nn\\
\label{octet_drbar}
&&~~~~~~~~~~~~~~~ \left. 
+ \frac{\hat X_t^4}{6} \,\left[1 + 3\,x_i\,(1 + \ln x_i) 
-  \ln \frac{m_{O_i}^2}{Q^2} \,+\, 6\,x_i^2\,f(x_i)~\right]~\right\}~,
\eea
where $x_i\equiv \msusy^2/m_{O_i}^2\,$, and the function $f(x)$ is defined as
\beq
\label{eq:fx}
f(x) ~=~ \frac{1}{1-4x}\,\left[\ln x + x \,\phi\left(\frac{1}{4x}\right)\right],
\eeq
$\phi(z)$ being the function defined in eq.~(45) of
ref.~\cite{Brignole:2001jy}. Special limits of the function in
eq.~(\ref{eq:fx}) above are $f(1/4) = -2\,(1+\ln4)/3\,$ and
$f(1) \approx -0.781302$. In the OS scheme the octet-scalar
contributions receive -- at the leading order in $\mt/\msusy$ -- the
shift
\beq
\label{octet_shift}
\delta\left(\Delta m_h^2\right)_{2\ell}^{O_i} ~=~
\frac{\alpha_s \mt^4}{\pi^3v^2}\,
\left\{ {\cal B}_i ~-~ \left(\hat X_t^2-\frac{\hat X_t^4}{6}\right)
\,\biggr[ 3\,{\cal B}_i ~+~ 2\,\ln\frac{m_{O_i}^2}{Q^2} ~-~ 2\biggr]
\,\right\}~,
\eeq
where
${\cal B}_i \equiv \hat B_0(\msusy^2,\msusy^2,m_{O_i}^2) =
-\ln(m_{O_i}^2/Q^2) + g(\msusy^2/m_{O_i}^2)\,$, with the function $g(x)$
defined as
\beq
\label{eq:gx}
g(x) ~=~ \left\{
\begin{array}{l}
2 - \left(1-\frac 1{2x}\right)\,\ln x 
- \frac 1x\,\sqrt{4\,x-1}\,\arctan\sqrt{4\,x-1}
~~~~~~~
(x \,>\, 1/4)
\\[3mm]
2 - \left(1-\frac 1{2x}\right)\,\ln x 
+ \frac 1x\,\sqrt{1-4\,x}\,\arctan{\rm \!h}\sqrt{1-4\,x}
~~~~~~
(x \,<\, 1/4)
\end{array}
\right. ~.
\eeq
Again, it can be easily checked that the explicit dependence on $Q$
cancels out in the sum of eqs.~(\ref{octet_drbar}) and
(\ref{octet_shift}).

\subsubsection{MRSSM with heavy Dirac gluino}
\label{sec:heavyglu}

The second simplified scenario we consider is the $R$-symmetric model
of section~\ref{sec:MRSSM}, in the limit of heavy Dirac gluino,
i.e.~$\mg \gg m_{\tilde t_i}\,$. This is a phenomenologically
interesting limit because Dirac gaugino masses are ``supersoft'',
i.e.~they can be substantially larger than the squark masses without
spoiling the naturalness of the model~\cite{Fox:2002bu}.

In the MRSSM the left and right stops do not mix, hence we set
$\theta_t = 0$ in our formulae, but we allow for the possibility of
different stop masses $m_{\tilde t_1}$ and $m_{\tilde t_2}$. In the
decoupling limit of the Higgs sector, where we neglect the mixing with
the heavy neutral states, the correction to the SM-like Higgs mass
reduces to
$~\Delta m_h^2 \,\approx\, \sin^2 \beta\, \left(\Delta
  \mathcal{M}_S^2\right)_{22}\,$. In analogy to eq.~(\ref{dmh_MS}),
the correction can in turn be decomposed in a dominant one-loop part,
a two-loop, MSSM-like $\oatas$ contribution and two-loop octet-scalar
contributions:
\beq
\label{dmh_R}
\Delta m_h^2 ~\approx~ \frac{3\,m^4_t}{8\,\pi^2 v^2}\,
\ln\frac{\tu\td}{m_t^4} 
~+~\left(\Delta m_h^2\right)_{2\ell}^{{\scriptscriptstyle {\rm ``MSSM"}}} 
+~ c^2_{\phi_{O}} \left(\Delta m_h^2\right)_{2\ell}^{O_1}
~+~ s^2_{\phi_{O}} \left(\Delta m_h^2\right)_{2\ell}^{O_2}~.
\eeq

Assuming that the top and stop masses in the one-loop part of the
correction are $\drbar$-renormalised parameters at the scale $Q$, and
expanding our results in inverse powers of $\g$, the contribution of
two-loop, MSSM-like diagrams involving gluons, gluinos or a four-stop
coupling reads
\bea
\left(\Delta m_h^2\right)_{2\ell}^{{\scriptscriptstyle {\rm ``MSSM"}}}\!&=& 
\frac{\as\,\mt^4}{4\,\pi^3v^2} 
\left\{~ \frac{2\,\g}{\tu}\left( 1 - \ln\frac{\g}{Q^2}~\right)
\,+\,\frac{2\pi^2}{3} - 2 - 6\,\ln\frac{\g}{\tu}+ 2\,\ln\frac{\t}{Q^2}
\right.\nn\\[2mm]
&& ~~~~~~~~~~~+
\frac{2\,\t}{\tu}\left(1- \ln\frac{\g}{Q^2}\right)
+ \ln^2\frac{\g}{\t} + \ln^2\frac{\g}{\tu} + 2\,\ln^2\frac{\t}{Q^2}
- 2\,\ln^2\frac{\tu}{Q^2}
\nn\\[2mm]
&& ~~~~~~~~~~~+\frac{2\,m_t^2}{\g} 
\left[  \frac{2\pi^2}{3}\left(2 + \frac{\tu}{m_t^2}\right) - 2 
- \left(8 + \frac{\t}{\tu}\right)\ln\frac{\g}{\t} 
- 4\,\ln\frac{\g}{\tu}
\right.
\nn\\[2mm]
&& ~~~~~~~~~~~~~~~~~~~~~~~-\left.
\frac{\tu}{\t}\left(2 + 6\,\ln \frac{\g}{\tu} +
\ln \frac{\g}{\t}\right)+2\left(2 + \frac{\tu}{\t}\right)
\ln \frac{\g}{\tu}\ln \frac{\g}{\t}~\right]\nn\\[2mm]
&&~~~~\,\left.
\phantom{\frac{\g}{\tu}}+~~ {\cal O}\left(\mg^{-4}\right)~\right\}
~+~~ \biggr[\tu ~\longrightarrow~ \td\biggr]~,
\label{mssmDR}
\eea
where the last term in square brackets represents the addition of
terms obtained from the previous ones by replacing $\tu$ with
$\td$. From eq.~(\ref{mssmDR}) above it is clear that, in the $\drbar$
scheme, the two-loop top-stop-gluino contributions to the SM-like
Higgs mass can become unphysically large when
$\mg \gg m_{\tilde t_i}\,$, due to the presence of terms enhanced by
$\g/\ti$. This non-decoupling behaviour of the corrections to the
Higgs mass in the $\drbar$ scheme has already been discussed in the
context of the MSSM in ref.~\cite{Degrassi:2001yf}. Indeed, the
correction in eq.~(\ref{mssmDR}) corresponds to the one obtained by
setting $\mu=A_t=0$ in the MSSM result. The terms enhanced by $\g/\ti$
can be removed by expressing the top and stop masses in the one-loop
part of the correction as OS parameters. After including the resulting
shifts in the two-loop correction, we find
\bea
\left(\Delta m_h^2\right)_{2\ell}^{{\scriptscriptstyle {\rm ``MSSM"}}}\!&=& 
\frac{\as\,\mt^4}{4\,\pi^3v^2} 
\left\{ \frac{2\pi^2}{3} - 1 - 6\, \ln \frac{\g}{m_t^2} 
- 3\, \ln^2 \frac{m_{\tilde{t}_1}^2}{m_t^2} 
+ 2 \,\ln^2 \frac{\g}{\tu}~\right. \nn\\[2mm]
&& ~~~~~~~~~~+\frac{m_t^2}{\g} 
\left[  \frac{4\pi^2}{3}\left(2 + \frac{\tu}{m_t^2}\right) - \frac{20}{3} 
- \frac{14 \,\tu}{3\,\mt^2} 
+ \frac{28}{3} \,\ln \frac{\tu}{\mt^2}\right.
\nn\\[2mm]
&& ~~~~~~~~~~~~~~~~~~~~ + \frac{2\,\tu}{\mt^2} 
\left(6 + \ln \frac{\tu}{\mt^2} + \ln \frac{\td}{\mt^2} \right)
\ln \frac{\tu}{\mt^2} 
+ \frac{\td}{\mt^2}\,\ln \frac{\tu}{\td} \nn\\[2mm]
&& ~~~~~~~~~~~~~~~~~~~~ -2\left( 12 
+ \frac{6\, \tu}{\mt^2} + 4\, \ln \frac{\tu}{\mt^2} 
+ \frac{3\,\tu + \td}{\mt^2} \ln \frac{\tu}{\mt^2}\right)
\ln \frac{\g}{\mt^2} \nn\\[2mm]
&& ~~~~~~~~~~~~~~~~~~~~ \left. \left.
\label{mssmOS}
+ 4 \left( 2 + \frac{\tu}{\mt^2}\right) \ln^2 \frac{\g}{\mt^2}~ \right] 
~+~ {\cal O}\left(\mg^{-4}\right)~\right\}
~+~~ \biggr[\tu ~\longleftrightarrow~ \td\biggr]~, \nn\\[2mm]
\eea
where the last term in square brackets represents the addition of
terms obtained from the previous ones by swapping $\tu$ and $\td$. By
taking the limit $m_{\tilde t_1} = m_{\tilde t_2} = m_{\tilde t}$ in
the equation above we recover eq.~(42) of ref.~\cite{Degrassi:2001yf}.

In the MRSSM, the contributions to $\Delta m_h^2$ arising from
two-loop diagrams with stops and octet scalars allow for fairly
compact expressions. If the stop masses in the one-loop part of the
correction are renormalised in the $\drbar$ scheme, those
contributions read
\beq
\left(\Delta m_h^2\right)_{2\ell}^{O_i} ~=~
-\frac{\alpha_s \mt^4}{2\,\pi^3v^2}\, \frac{\g}{\tu}\,
\left\{ \,1 - \ln \frac{\tu}{Q^2} + f\left(\frac{\tu}{\oi}\right)\right\}
~~+~~ \biggr[\tu ~\longrightarrow~ \td\biggr]~,
\label{Roctet_drbar}
\eeq
where $f(x)$ is the function defined in eq.~(\ref{eq:fx}). For OS stop
masses, the octet-scalar contributions to $\Delta m_h^2$ read
instead
\beq
\left(\Delta m_h^2\right)_{2\ell}^{O_i} ~=~
-\frac{\alpha_s \mt^4}{2\,\pi^3v^2}\, \frac{\g}{\tu}\,
\left\{ \,1 - \ln \frac{\tu}{\oi} + f\left(\frac{\tu}{\oi}\right)
- g\left(\frac{\tu}{\oi}\right)\right\}
~~+~~ \biggr[\tu ~\longrightarrow~ \td\biggr]~,
\label{Roctet_OS}
\eeq
where $g(x)$ is the function defined in eq.~(\ref{eq:gx}). It would
appear from eqs.~(\ref{Roctet_drbar}) and (\ref{Roctet_OS}) above
that, independently of the renormalisation scheme adopted for the stop
masses, the octet-scalar contributions to $\Delta m_h^2$ are enhanced
by a factor $\g$. This is due to the fact that the trilinear
squark-octet interaction, see eq.~(\ref{eq:interaction}), is
proportional to the Dirac mass term $m_D$ -- i.e., to
$\mg\,$. However, as discussed in section~\ref{sec:adjoints}, one of
the mass eigenvalues for the octet scalars -- to fix the notation, let
us assume it is $m^2_{O_1}$ -- does in turn grow with the gluino mass,
namely $m^2_{O_1} \approx 4\,m_D^2$ when $m_D^2$ becomes much larger
than the soft SUSY-breaking mass terms for the octet
scalars. Expanding the corresponding contribution to $\Delta m_h^2$ in
inverse powers of $m^2_{O_1}$ we find, in the $\drbar$ scheme,
\bea
\left(\Delta m_h^2\right)_{2\ell}^{O_1}\!&=& 
-\frac{\as\,\mt^4}{4\,\pi^3v^2}\, \frac{\g}{\os}
\left\{\, 2\,\frac{\os}{\tu}\left(1-\ln\frac{\os}{Q^2}\right)\,+\,
\frac{2\pi^2}{3} \,+ \,8\, \ln \frac{\tu}{\os} 
\,+\, 2\, \ln^2 \frac{\tu}{\os} \right. \nn\\[2mm]
&& ~~~~~~~~~~~~~~~~~~~~+ \left. \frac{4\,\tu}{\os} 
\left[\pi^2 - 2 + 10\,\ln\frac{\tu}{\os} 
+ 3\,\ln^2\frac{\tu}{\os}
~ \right] 
~+~ {\cal O}\left(m_{O_1}^{-4}\right)~\right\}\nn\\[2mm]
&+& \biggr[\tu ~\longrightarrow~ \td\biggr]~,
\label{octet_DR_exp}
\eea
which does indeed contain potentially large terms enhanced by the
ratio $\g/\ti$. Note that those terms cancel only partially the
corresponding terms in the MSSM-like contribution -- see the first term
in the curly brackets of eq.~(\ref{mssmDR}) -- leaving residues
proportional to $\,\g/\ti\,\ln(\os/\g)\,$. On the other hand, in the
OS scheme we find 
\bea
\left(\Delta m_h^2\right)_{2\ell}^{O_1}\!&=& 
-\frac{\as\,\mt^4}{4\,\pi^3v^2}\, \frac{\g}{\os}
\left\{\, \frac{2\pi^2}{3} \,-\, 1 \,+ \,6\, \ln \frac{\tu}{\os} 
\,+\, 2\, \ln^2 \frac{\tu}{\os} \right. \nn\\[2mm]
&& ~~~~~~~~~~~~~~~~~~~~+ \left. \frac{4\,\tu}{\os} 
\left[\pi^2 - \frac{17}{6} + 9\,\ln\frac{\tu}{\os} 
+ 3\,\ln^2\frac{\tu}{\os}
~ \right] 
~+~ {\cal O}\left(m_{O_1}^{-4}\right)~\right\}\nn\\[2mm]
&+& \biggr[\tu ~\longrightarrow~ \td\biggr]~.
\label{octet_OS_exp}
\eea
Thus, we see that in the OS scheme the contribution to $\Delta m_h^2$
from two-loop diagrams involving the heaviest octet scalar $O_1$ does
not grow unphysically large when $\g$ increases, because the ratio
$\g/\os$ tends to $1/4$.  In contrast, for the contribution of the
lightest octet scalar $O_2$, whose squared mass does not grow with
$\g$, the unexpanded formulae in eqs.~(\ref{Roctet_drbar}) and
(\ref{Roctet_OS}) should always be used. However, in the total
correction to $m_h^2$ -- see eq.~(\ref{dmh_R}) -- the $\g$ enhancement
of $\left(\Delta m_h^2\right)_{2\ell}^{O_2}$ is compensated for by the
factor $s^2_{\phi_O}$, which, as discussed in
section~\ref{sec:adjoints}, is in fact suppressed by $\mg^{-4}$ in the
heavy-gluino limit. In summary, we find that, in the MRSSM with heavy
Dirac gluino, neither of the octet scalars can induce unphysically
large contributions to $\Delta m_h^2$, as long as the stop masses in
the one-loop part of the correction are renormalised in the OS scheme.


\section{Numerical examples}
\label{sec:numerical}

In this section we discuss the numerical impact of the two-loop
$\oatas$ corrections to the Higgs boson masses whose computation was
described in the previous section. As we did for the simplified
formulae of section~\ref{sec:simpli}, we focus on ``decoupling''
scenarios in which the lightest neutral scalar is SM-like and the
superpotential couplings $\lambda_{S,T}$ are subdominant with respect
to the top Yukawa coupling. Our purpose here is to elucidate the
dependence of the corrections to the SM-like Higgs boson mass $m_h$ on
relevant parameters such as the stop masses and mixing and the gluino
masses, rather than provide accurate predictions for all Higgs boson
masses in realistic scenarios. We therefore approximate the one-loop
part of the corrections with the dominant top/stop contributions at
vanishing external momentum, obtained by combining the formulae for
the Higgs mass matrices given for MDGSSM and MRSSM in
sections~\ref{sec:MDGSSM} and \ref{sec:MRSSM}, respectively, with the
one-loop functions given in eq.~(\ref{eq:F1loop}).  We recall that a
computation of the Higgs boson masses in models with Dirac gauginos
could also be obtained in an automated way by means of the package
{\tt SARAH}~\cite{Staub:2008uz, Staub:2009bi, Staub:2010jh,
  Staub:2012pb, Staub:2013tta, Staub:2015kfa}. That would include the
full one-loop corrections~\cite{Staub:2012pb} and the two-loop
corrections computed in the gaugeless limit at vanishing external
momentum~\cite{Goodsell:2014bna, Goodsell:2015ira}. However, the
computation implemented in {\tt SARAH} employs the $\drbar$
renormalisation scheme, and does not easily lend itself to an
adaptation to the OS scheme which, as discussed in
section~\ref{sec:heavyglu}, can be more appropriate in scenarios with
heavy gluinos.

The SM parameters entering our computation of the Higgs boson masses,
which we take from ref.~\cite{Agashe:2014kda}, are the $Z$ boson mass
$\MZ = 91.1876$~GeV, the Fermi constant
$G_F = 1.16637\times10^{-5}$~GeV$^{-2}$ (from which we extract
$v = (2 \,\sqrt 2 \,G_F)^{-1/2} \approx 174$~GeV), the pole top-quark
mass $\mt = 173.21$~GeV and the strong gauge coupling of the SM in the
$\msbar$ renormalisation scheme, $\overline
{\alpha}_s(\MZ)=0.1185$. Concerning the SUSY parameters entering the
scalar mass matrix at tree-level, we set $\lambda_S=\lambda_T=0$ and
push the parameters that determine the heavy-scalar masses to
multi-TeV values, so that
$(m_h^2)^{\rm tree} \approx \MZ^2\,\cos^22\beta$. We also set
$\tb=10$, so that the tree-level mass of the SM-like Higgs boson is
almost maximal but the corrections from diagrams involving sbottom
squarks, which we neglect, are not particularly enhanced. For the
parameters in the stop mass matrices we take degenerate soft
SUSY-breaking masses $m_Q=m_U =\MS$, we neglect D-term-induced
electroweak contributions and we treat the whole left-right mixing
term $X_t = A_t - \mu\cot\beta$ as a single input. Finally, for what
concerns the parameters that determine the gluino and octet-scalar
masses we focus again on the case of purely-Dirac gluinos, with
$m_{\tilde g_1} = -m_{\tilde g_2} = \mg\,$ and
$R_{11}^2=R_{12}^2=1/2\,$. We also take a vanishing soft SUSY-breaking
bilinear $B_O$, so that $\phi_O=0$ and only the CP-even octet scalar
$O_1$, with mass $m^2_{O_1} = m_O^2 + 4\,\mg^2\,$, participates in the
$\oatas$ corrections to the Higgs masses.

\subsection{An example in the MDGSSM}

\begin{figure}[p]
\begin{center}
\includegraphics[width=0.65\textwidth]{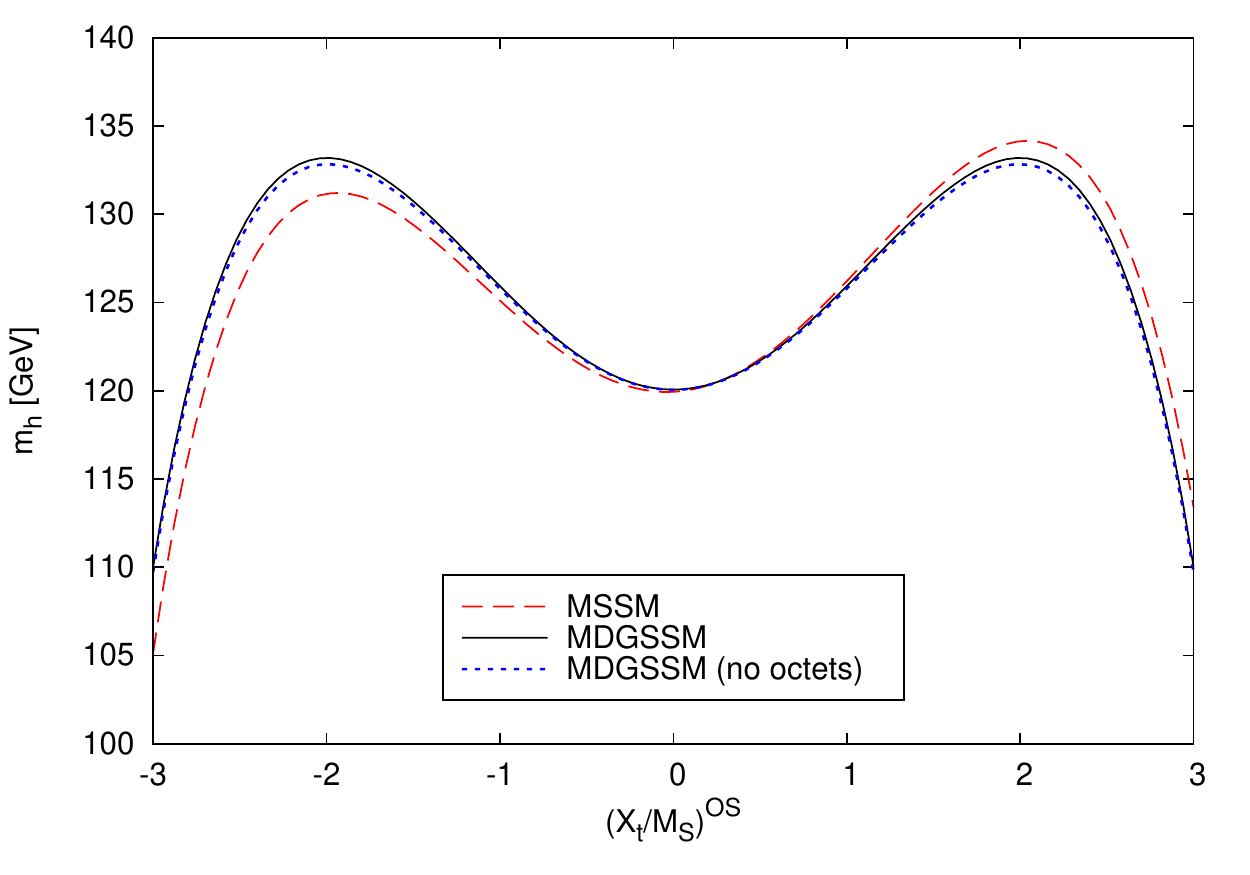}
\end{center}
\vspace*{-3mm}
\caption{Mass of the SM-like Higgs boson as a function of
  $(X_t/\MS)^{\rm OS}$, for $\tan\beta=10$, $\MS=1.5$~TeV and
  $\mg=m_O=2$~TeV. The dashed curve represents the MSSM result, whereas
  the solid (dotted) curve represents the MDGSSM result with (without)
  the octet-scalar contributions.}
\label{fig:models}
\end{figure}

In figure \ref{fig:models} we illustrate some differences between the
$\oatas$ corrections to the SM-like Higgs boson mass in the MDGSSM and
in the MSSM. We plot $m_h$ as a function of the ratio $X_t/\MS$,
setting $\MS=1.5$~TeV and $\mg=m_O=2$~TeV and adopting the OS
renormalisation scheme for the parameters $\mt$, $\MS$ and $X_t$. We
employ the renormalisation-group equations of the SM to evolve the
coupling $\overline{\alpha}_s$ from the input scale $\MZ$ to the scale
$\MS$, then we convert it to the $\drbar$-renormalised coupling of the
considered SUSY model, which we denote as $\hat{\alpha}_s(\MS)$, by
including the appropriate threshold corrections (in this step, we
assume that all soft SUSY-breaking squark masses are equal to $\MS$).
The solid (black) and dashed (red) curves in figure~\ref{fig:models}
represent the SM-like Higgs boson mass in the MDGSSM and in the MSSM,
respectively. The comparison between the two curves highlights the
fact that, in contrast with the case of the MSSM, in the MDGSSM with
purely-Dirac gluinos the $\oatas$ corrections to $m_h$ are symmetric
with respect to a change of sign in $X_t$. As mentioned in
section~\ref{sec:commonMS}, this stems from cancellations between
terms proportional to odd powers of the gluino masses. In the points
where $m_h$ is maximal, which in the OS calculation happens for
$|X_t/\MS| \approx 2$, the difference between the MDGSSM and MSSM
predictions for $m_h$ is about $1$ or $2$~GeV, depending on the sign
of $X_t$. Finally, the dotted (blue) curve in figure \ref{fig:models}
represents the prediction for $m_h$ obtained in the MDGSSM by omitting
the contributions of two-loop diagrams involving the octet
scalars. The comparison between the solid and dotted curves shows
that, in the considered point of the parameter space, the effect on
$m_h$ of the octet-scalar contributions is positive but rather small,
of the order of a few hundred MeV. Varying the parameters $\MS$, $\mg$
and $m_O$ by factors of order two around the values used in
figure~\ref{fig:models}, we find that this is a typical size
for the octet-scalar contributions to $m_h$ in the OS scheme.

\begin{figure}[p]
\begin{center}
\includegraphics[width=0.65\textwidth]{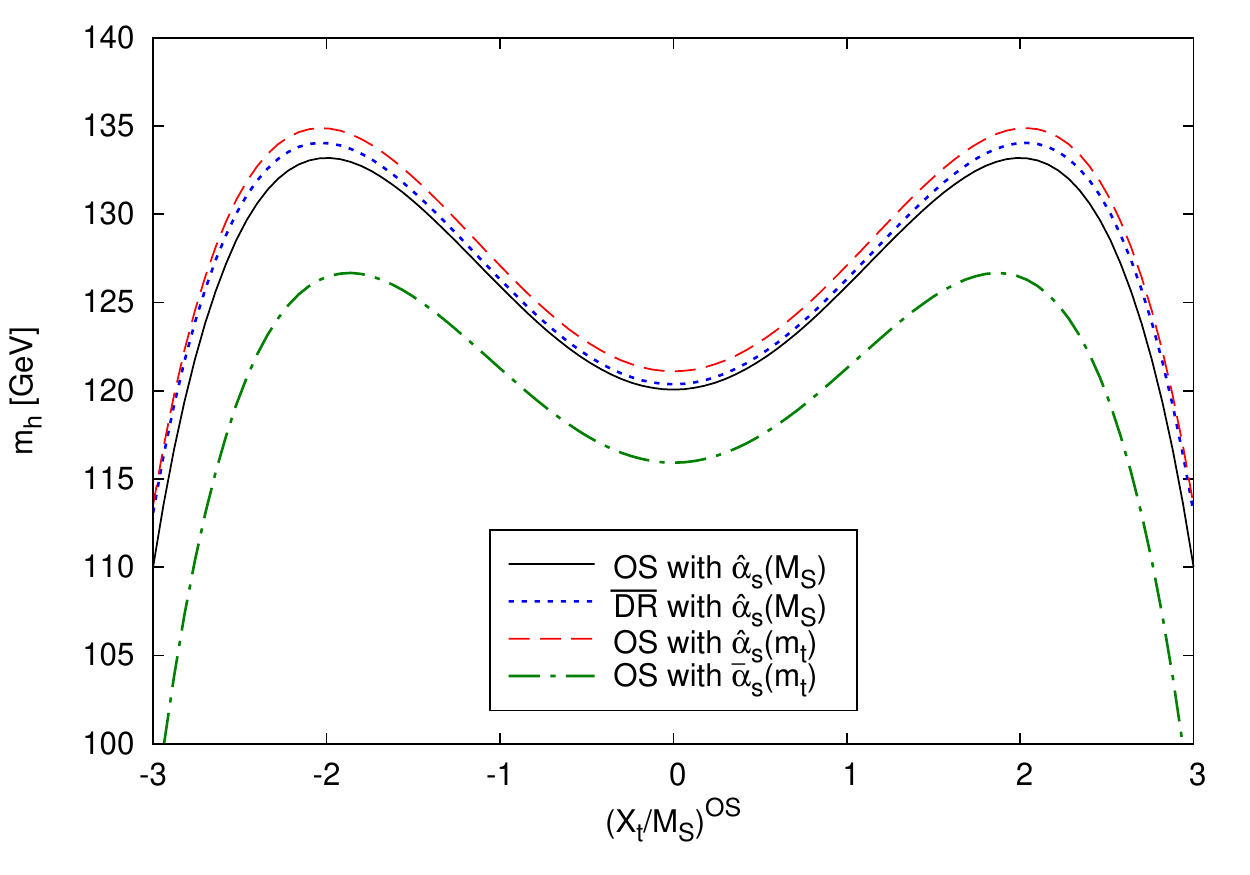}
\end{center}
\vspace*{-3mm}
\caption{Different determinations of the SM-like Higgs boson mass in
  the MDGSSM as a function of $(X_t/\MS)^{\rm OS}$, for the same
  choices of parameters as in figure~\ref{fig:models}. The solid curve
  represents the original OS calculation; the dotted curve represents
  the $\drbar$ calculation; the dashed and dot-dashed curves were
  obtained using $\hat{\alpha}_s(\mt)$ and $\overline{\alpha}_s(\mt)$,
  respectively, in the OS calculation instead of
  $\hat{\alpha}_s(\MS)$.}
\label{fig:uncert}
\end{figure}

A discussion of the theoretical uncertainty of our calculation is now
in order. In our numerical examples we are not implementing the
full one-loop corrections to the Higgs boson masses, nor the two-loop
corrections beyond $\oatas$ that are available in {\tt SARAH}, in order
to focus purely on the ${\cal O}(\at\as)$ corrections. Therefore
the only sources of uncertainty that we can meaningfully estimate are the
uncomputed effects of ${\cal O}(\at\as^2)$, i.e.~those arising from
genuine three-loop diagrams with four strong-interaction vertices and
from SUSY-QCD renormalisation effects of the parameters entering the
one- and two-loop corrections. A common procedure for estimating those
effects consists in comparing the results of the $\oatas$ calculation
of $m_h$ in the OS scheme with the results obtained by $i)$ converting
the OS input parameters -- i.e., the top mass and the stop masses and
mixing -- to the $\drbar$ scheme by means of $\oas$ shifts, and $ii)$
computing $m_h$ using these $\drbar$ parameters in both the one-loop
and two-loop corrections, with the appropriate $\drbar$ formulae for
the $\oatas$ corrections. The two sources of ${\cal O}(\at\as^2)$
discrepancies in such a comparison are the omission of terms quadratic
in $\delta x_k$ in the expansion of the one-loop part of the
corrections, eq.~(\ref{2lshifts}), and the different definition of the
top and stop parameters entering the two-loop part of the
corrections. In figure~\ref{fig:uncert} we illustrate the
renormalisation-scheme dependence of the $\oatas$ determination of
$m_h$, in the same MDGSSM scenario as in figure~\ref{fig:models}. The
solid (black) curve represents the results of the original OS
calculation, whereas the dotted (blue) curve represents the results of
the $\drbar$ calculation described above (note that both curves are
plotted as functions of the ratio of OS parameters $X_t/\MS$). The
comparison between the solid and dotted curves would suggest a rather
small impact of the uncomputed ${\cal O}(\at\as^2)$ corrections, of
the order of one GeV or even less (at least for the considered
scenario).

Besides the top mass and the stop masses and mixing, there are a few
more parameters entering the $\oatas$ corrections to the Higgs boson
masses whose $\oas$ definition amounts to a three-loop ${\cal
  O}(\at\as^2)$ effect, namely the gluino and octet-scalar masses and
the strong gauge coupling itself. Concerning the masses, in an OS
calculation it seems natural to interpret them as pole ones. For
$\as$, on the other hand, there is no obvious ``on-shell'' definition
available, and different choices of scheme, scale and even underlying
theory -- while all formally equivalent at $\oatas$ for the Higgs-mass
calculation -- can lead to significant variations in the numerical
results. As mentioned earlier, the solid curve in
figure~\ref{fig:uncert} was obtained with top/stop parameters in the
OS scheme, but with $\alpha_s$ defined as the $\drbar$-renormalised
coupling of the MDGSSM at the stop-mass scale,
i.e.~$\hat{\alpha}_s(\MS)$. However, since both stop squarks and top
quarks enter the relevant two-loop diagrams, it would not seem
unreasonable to evaluate the strong gauge coupling at the top-mass
scale either. The dashed (red) and dot-dashed (green) curves in
figure~\ref{fig:uncert} represent the predictions for $m_h$ obtained
with top/stop parameters still in the OS scheme, but with $\alpha_s$
defined as the $\drbar$-renormalised coupling of the MDGSSM at the
top-mass scale, $\hat{\alpha}_s(\mt)$, and as the
$\msbar$-renormalised coupling of the SM at the same scale,
$\overline{\alpha}_s(\mt)$, respectively. The comparison of these two
curves with the solid curve shows that a variation in the definition
of the coupling $\alpha_s$ entering the two-loop corrections provides
a less-optimistic estimate of the uncertainty associated to the ${\cal
  O}(\at\as^2)$ corrections compared with the scheme variation of the
top/stop parameters. In particular, for the considered scenario the
use of $\overline{\alpha}_s(\mt)$ would induce a negative variation
with respect to the results obtained with $\hat{\alpha}_s(\MS)$ of
about $4$~GeV for $X_t \approx 0$ and about $7$~GeV for $|X_t/\MS|
\approx 2$. In contrast, the use of $\hat{\alpha}_s(\mt)$ would induce
a positive variation of about $1$~GeV for $X_t \approx 0$ and about
$2$~GeV for $|X_t/\MS| \approx 2$, i.e.~more modest than the previous
one but still larger than the one induced by a scheme change in the
top/stop parameters. While remaining agnostic about the true size (and
sign) of the three-loop ${\cal O}(\at\as^2)$ corrections, we take this
as a cautionary tale against putting too much stock in any single
estimate of the theoretical uncertainty of a fixed-order calculation
of $m_h$ in scenarios with TeV-scale superparticles.

\subsection{An example in the MRSSM}

In our second numerical example we consider the MRSSM, and illustrate
the dependence of the SM-like Higgs boson mass on the gluino mass. In
ref.~\cite{Diessner:2015yna} it was pointed out that, for multi-TeV
values of $\mg$, the contribution of two-loop diagrams involving octet
scalars can increase the prediction for $m_h$ by more than
$10$~GeV. We will show that such large effects are related to the
non-decoupling behaviour of the $\drbar$ calculation of $m_h$ that we
discussed in section~\ref{sec:heavyglu}, and that the octet-scalar
contributions are much more modest in an OS calculation.

\begin{figure}[t]
\begin{center}
\includegraphics[width=0.65\textwidth]{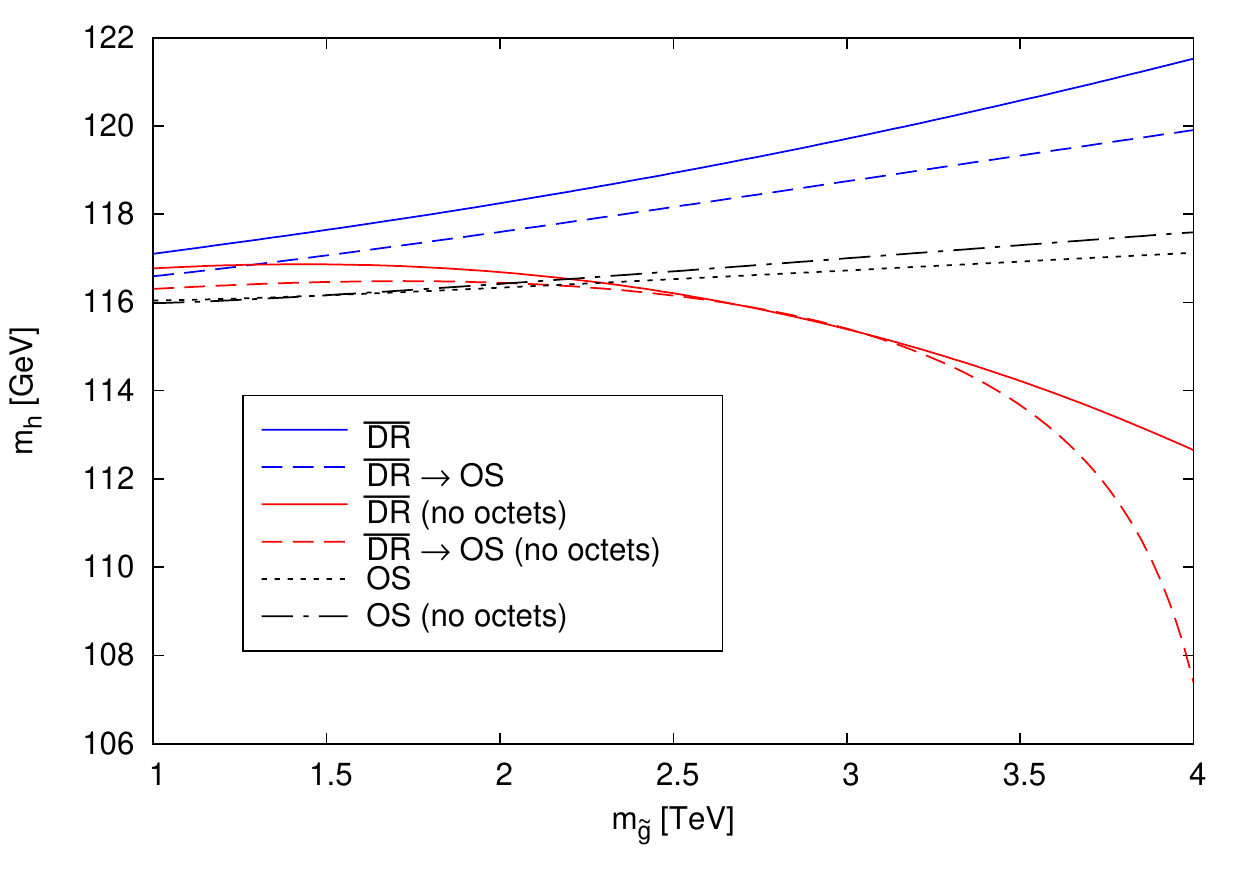}
\end{center}
\vspace*{-3mm}
\caption{Mass of the SM-like Higgs boson as a function of $\mg$ in the
  MRSSM, for $\tan\beta=10$, $\MS=1$~TeV and $m_O=2$~TeV. The meaning
  of the different curves is explained in the text.}
\label{fig:mrssm}
\end{figure}

The upper (blue) and lower (red) solid curves in
figure~\ref{fig:mrssm} represent the SM-like Higgs boson mass obtained
from the $\drbar$ calculation as a function of $\mg$, with and without
the octet-scalar contributions, respectively. We set $m_O = 2$~TeV and
$\MS=1$~TeV. The latter is interpreted as a $\drbar$-renormalised soft
SUSY-breaking parameter evaluated at a scale equal to $\MS$ itself,
which means that each point in the solid curves corresponds to a
different value of the physical stop masses. Both curves show a marked
dependence on $\mg$, and the comparison between them shows that, for
the highest value of $\mg$ considered in the plot, the effect on $m_h$
of the two-loop octet-scalar contributions does indeed grow to about
$9$~GeV. However, as can be seen in the explicit formulae for the
two-loop corrections in the $\drbar$ scheme of eqs.~(\ref{mssmDR}) and
(\ref{octet_DR_exp}), this marked dependence of both the gluino and
octet-scalar contributions on $\mg$ is induced by terms enhanced by
the ratio $\mg^2/\MS^2$. When that ratio becomes large, which in
Dirac-gaugino models can occur naturally, the size of the two-loop
$\oatas$ corrections to $m_h$ can grow up to a point where the
accuracy of the perturbative expansion is called into question.
To visualise this aspect, we perform a change of renormalisation
scheme for the top and stop masses that mirrors the one represented by
the dotted curve in figure~\ref{fig:uncert}. The upper (blue) and
lower (red) dashed curves in figure~\ref{fig:mrssm} represent the
values of $m_h$ obtained with and without octet-scalar contributions,
respectively, after converting the $\drbar$ stop masses into the
physical ones and using the latter, together with the physical top
mass, in both the one-loop and two-loop corrections, with the
appropriate OS formulae for the $\oatas$ corrections. For our choice
of the $\drbar$ input parameter $\MS(\MS)=1$~TeV, we find that the
physical stop masses range between $1072$~GeV and $1392$~GeV for the
values of $\mg$ shown in the plot. If the octet-scalar contributions
to the $\oas$ stop self-energies are omitted, the stop masses range
instead between $1049$~GeV and $346$~GeV, i.e.~they become smaller for
increasing $\mg$ (indeed, in this case $\mg$ cannot be pushed to
values much larger than those shown in the plot without rendering the
stop masses tachyonic). The comparison between the solid and dashed
curves shows that the scheme dependence of the $\oatas$ calculation of
$m_h$ becomes increasingly worse at large values of $\mg$, especially
in the lower curves where the octet-scalar contributions are omitted.
Finally, the (black) dotted and dot-dashed curves in
figure~\ref{fig:mrssm} represent the predictions for $m_h$ obtained
directly from the OS calculation with and without octet-scalar
contributions, respectively. In this case the input $\MS=1$~TeV is
interpreted as an OS-renormalised parameter, meaning that the physical
stop masses correspond to $(\MS^2 + \mt^2)^{1/2}\approx1015$~GeV for
all points in the curves. We stress that direct comparisons between
these two curves and the solid (and dashed) ones would not be
appropriate, because they refer to different points of the MRSSM
parameter space. However, the dotted and dot-dashed curves show that,
when the physical stop masses are taken as input, the prediction for
$m_h$ in the MRSSM depends only mildly on the value of $\mg$, and the
effect of the octet-scalar contributions is below one GeV. This is
explained by the fact that, as discussed in
section~\ref{sec:heavyglu}, in the OS scheme there are no terms
enhanced by $\mg^2/\MS^2$ in either the gluino or the octet-scalar
contributions to the $\oatas$ corrections.

\begin{figure}[t]
\begin{center}
\includegraphics[width=0.65\textwidth]{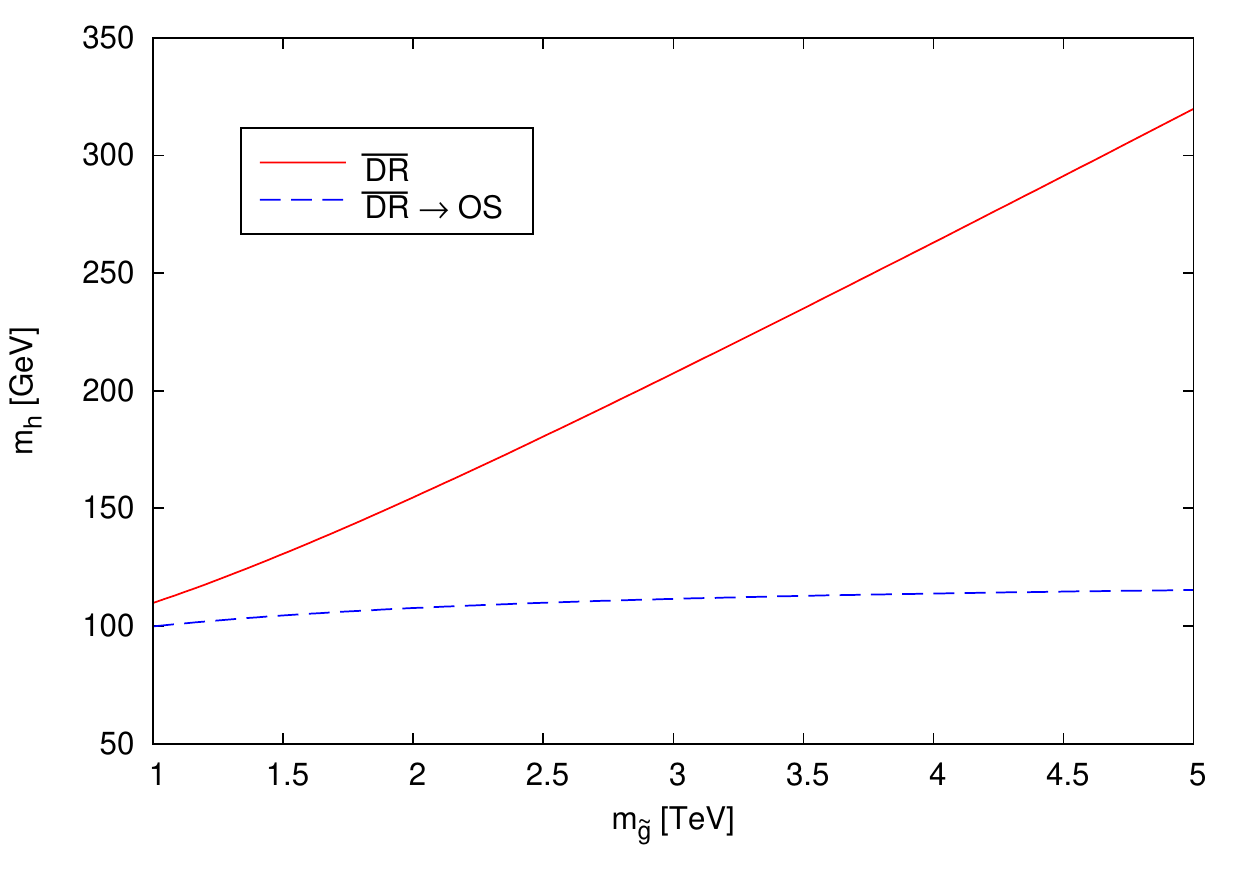}
\end{center}
\vspace*{-3mm}
\caption{Mass of the SM-like Higgs boson as a function of $\mg$ in the
  {\em supersoft} limit of the MRSSM, for $\tan\beta=10$. The solid
  curve represents the results of the $\drbar$ calculation, in which the
  two-loop $\oatas$ corrections become unphysically large. The dashed
  curve was obtained by converting the top and stop masses to the OS
  scheme and using the corresponding formulae for the $\oatas$
  corrections.}
\label{fig:supersoft}
\end{figure}

Before concluding, we note that there are extreme situations in which
a $\drbar$ calculation of $m_h$ is not workable at all, and a
conversion to the OS scheme such as the one represented by the dashed
lines in figure~\ref{fig:mrssm} is necessary. In the so-called {\em
  supersoft} scenario, all soft SUSY-breaking masses vanish, and
sizeable sfermion masses -- proportional to the Dirac-gaugino masses
-- are induced only by radiative corrections. Such a scenario can be
realised e.g.~in the MRSSM by setting $m_O=0$ and $\MS=0$, where the
latter is interpreted as a $\drbar$-renormalised parameter. At the
scale where this condition is imposed, the $\drbar$ stop masses
coincide with the top mass, with the result that, in the $\drbar$
calculation, the one-loop correction in the first term of
eq.~(\ref{dmh_R}) vanishes, while the two-loop corrections in
eqs.~(\ref{mssmDR}) and (\ref{octet_DR_exp}) contain terms enhanced by
$\mg^2/\mt^2$ (concerning the octet-scalar contributions, we recall
that $m_{O_1} = 2\,\mg$ in this scenario). Since the Dirac-gluino mass
needs to be in the multi-TeV range to generate realistic values for
the physical stop masses, the non-decoupling terms in the two-loop
corrections can become unphysically large. This is illustrated by the
solid (red) curve in figure~\ref{fig:supersoft}, which represents the
SM-like Higgs boson mass obtained with the $\drbar$ calculation as a
function of the gluino mass (here we fix the renormalisation scale as
$Q=\mt$ and use $\overline \alpha_s(\mt)$ in the two-loop
corrections). It appears that the $\drbar$ prediction for $m_h$
becomes essentially proportional to $\mg$, and quickly grows to
nonsensical values as the latter increases. In contrast, the dashed
(blue) curve is obtained with the same procedure as the dashed curves
in figure~\ref{fig:mrssm}, i.e.~by computing the physical stop masses
at $\oas$ as a function of $\mg$ and using them in conjunction with the
appropriate OS formulae for the $\oatas$ corrections to $m_h$. In our
example the stop masses range between $302$~GeV and $1272$~GeV, while
the SM-like Higgs boson mass shows only a mild dependence on $\mg$ and
remains confined to values well below the observed one.

\section{Conclusions}
\label{sec:conclusion}

Supersymmetric models with Dirac gaugino masses have attracted
considerable attention in the past few years, because they are subject
to looser experimental constraints and require less fine-tuning than
the MSSM. Besides the extended gaugino sector, such models feature
additional colourless scalars which mix with the usual Higgs doublets
of the MSSM, as well as additional coloured scalars in the octet
representation of $SU(3)$ which contribute to the Higgs boson masses
at the two-loop level. In this paper we presented a computation of the
dominant two-loop corrections to the Higgs boson masses in
Dirac-gaugino models, relying on effective-potential techniques that
had previously been applied to the MSSM~\cite{Degrassi:2001yf} and to
the NMSSM~\cite{Degrassi:2009yq}. We obtained analytic formulae for
the $\oatas$ corrections to the scalar and pseudoscalar Higgs mass
matrices valid for arbitrary choices of parameters in the squark and
gaugino sectors, both in the $\drbar$ and in the OS renormalisation
schemes, which we make available upon request as a fortran code. We
also presented compact approximate formulae for the dominant
corrections to the mass of the SM-like Higgs boson, valid under a
number of simplifying assumptions for the SUSY parameters. Finally, we
studied the numerical impact of the newly-computed corrections on the
predictions for the SM-like Higgs boson mass in some representative
scenarios. In particular, we elucidated the differences between the
predictions for $m_h$ in the MSSM and those in its Dirac-gaugino
extensions; we discussed the theoretical uncertainty of our
predictions stemming from uncomputed higher-order corrections; we
stressed that a judicious choice of renormalisation scheme is required
to obtain reliable predictions in scenarios where the gluinos are much
heavier than the squarks, which can occur naturally in Dirac-gaugino
models. If our community's hopes are fulfilled and the run II of the
LHC brings on a wealth of new discoveries, our results will contribute
to their accurate interpretation in the framework of a well-motivated
SUSY extension of the SM.

\vfill
\newpage

\section*{Acknowledgements}

M.~D.~G.~thanks Florian Staub for collaboration on related topics.
This work was supported in part by French state funds managed by the
Agence Nationale de la Recherche (ANR), in the context of the LABEX
ILP (ANR-11-IDEX-0004-02, ANR-10-LABX-63). J.~B.~was supported by a
scholarship from the Fondation CFM. M.~D.~G.~and P.~S.~acknowledge
support from the ANR grant ``HiggsAutomator''
(ANR-15-CE31-0002). P.~S.~was supported in part by the Research
Executive Agency (REA) of the European Commission under the Initial
Training Network ``HiggsTools'' (PITN-GA-2012-316704), and by the
European Research Council (ERC) under the Advanced Grant ``Higgs@LHC''
(ERC-2012-ADG\_20120216-321133).


\appendix
\label{sec:appendix}

\section{Derivatives of the two-loop effective potential}
\label{sec:2lderivs}
\allowdisplaybreaks

We present here the derivatives of the two-loop effective potential
used to calculate the Higgs masses in section \ref{sec:2loopcalc}. 
We recall that the effective potential and its derivatives are
expressed in units of $\as\,C_F N_c/(4\pi)^3$.
The derivatives of the first term in eq.~(\ref{V2loop}) can be
trivially obtained by multiplying the formulae in appendix C of
ref.~\cite{Degrassi:2009yq} by $R_{1i}^2$ and summing over the two
gluino masses $\mgi\,$, hence we do not repeat them here. The only
exception is the single derivative of $\Delta V^{\,\as}_{\smallMSSM}$
with respect to $\t$, which was not needed in
ref.~\cite{Degrassi:2009yq}. Adapted to the Dirac-gaugino case, it
reads
\beq
\frac{~\,\partial \Delta V^{\,\as}}{\partial \t} ~= ~
\sum_{i=1}^2 \,R_{1i}^2 \, \frac{~\partial \Delta V_{\tilde g_i}}{\partial \t}~,
\eeq
with
\bea
\frac{\partial \Delta V_{\tilde g_i}}{\partial \t} \!\!&=& 
2\,\t\left(3-4\,\ln\frac{\t}{Q^2}+3\,\ln^2\frac{\t}{Q^2}\right)
~+~
2\,\left[\t\,\ln\frac{\gi}{\t}
+\tu\biggr(2-\ln\frac{\t}{Q^2}-\ln\frac{\gi}{Q^2}\biggr)
\right]\,\ln\frac{\tu}{Q^2}\nn\\[3mm]
&+&\!\!\!
\left[2\,(\tu-\t)-\frac{\mgi\,\tu\sdt}{\mt}\right]\ln\frac{\t}{Q^2}
\ln\frac{\gi}{Q^2} - 2\,\gi\biggr(3-2\,\ln\frac{\gi}{Q^2}\biggr)
-\tu\!\left(4-\frac{5\,\mgi\sdt}{\mt}\right)\nn\\[3mm]
&-&\frac{\mgi\,\sdt}{\mt}\,\left[
 \left(3\,\t-\gi\right)\,\ln\frac{\gi}{\t}
\,+\,\tu\biggr(4-\ln\frac{\t}{Q^2}-\ln\frac{\gi}{Q^2}\biggr)
\right]\,\ln\frac{\tu}{Q^2}\nn\\[3mm]
&+&\!2 \left[
\frac{\gi}{\t}\,(\gi-\t-\tu) - \frac{\Delta_{\tilde g_i}}{\t}
\,+\,\frac{\mgi\,\sdt}{\mt}\,\biggr(
\t-\gi-\tu + \frac{\Delta_{\tilde g_i}}{2\,\t}\biggr)\right]\,
\Phi(\tu,\gi,\t)\nn\\[3mm]
&+&~\biggr[\tu\,\rightarrow\,\td~,~~~~\sdt\,\rightarrow\,-\sdt \biggr]~,
\eea
where $Q$ is the renormalisation scale, the function $\Phi(x,y,z)$ is
defined in appendix D of ref.~\cite{Degrassi:2009yq}, and we used the
shortcut
\beq
\Delta_{\tilde g_i} ~\equiv~  (\gi - \t - \tu)^2 - 4 \, \t \tu~.
\eeq
The derivatives of the octet-scalar contribution $\Delta V_{O_i}$, computed
at the minimum of the potential, are
\bea
\frac{\partial\Delta V_{O_i}}{\partial \cdtt^2} &=&  
-2\bigg[I (\tu, \tu,\oi) + I (\td, \td,\oi) - 2 I (\tu, \td,\oi)\bigg]~,\\[3mm]
\label{eq:dvdstu}
\frac{\partial\Delta V_{O_i}}{\partial \tu} &=& 
2\, \Big( \ln \frac{\tu}{Q^2} - 1 \Big)^2 
+ 2\, \sdt^2 \ln \frac{\oi}{\tu} \ln\frac{\tu}{\td} \nn\\
&& - 2 \,\bigg[ \cdt^2 \Phi (\tu, \tu, \oi) +\sdt^2 
\frac{\oi - \tu + \td}{\oi} \Phi(\tu, \td,\oi) \bigg]~,\\[7mm]
\frac{\partial^2\Delta V_{O_i}}{(\partial \tu)^2} &=&
-  \frac{4 \,\sdt^2}{\tu} + \frac{ 4 \,\cdt^2}{\oi - 4 \tu} 
\bigg[ \frac{\oi}{\tu} \Big(\ln \frac{\oi}{Q^2} - 1\Big) 
- 4 \Big(\ln \frac{\tu}{Q^2} - 1\Big) - \Phi(\tu,\tu,\oi)\bigg] \nn\\
&&  +  \frac{ 4 \,\sdt^2}{\Delta_{O_i}}\,
\bigg[ \frac{\oi}{\tu} (\oi -\tu -\td)  \ln\frac{\oi}{Q^2}
- (\oi -\tu +\td)  \ln\frac{\tu}{Q^2}\nn\\
&&  
~~~~~~~~~~~~ - \frac{\td}{\tu} (\oi +\tu -\td)  \ln\frac{\td}{Q^2} 
- 2\, \td\, \Phi(\tu, \td,\oi)\bigg]~,\\[3mm]
\label{eq:dvdcdtdstu}
\frac{\partial^2\Delta V_{O_i}}{ \partial \tu \partial \cdtt^2} &=&
-2 \bigg[ \ln \frac{\oi}{\tu} \ln \frac{\tu}{\td} + \Phi (\tu, \tu, \oi)
- \frac{\oi - \tu + \td}{\oi}\, \Phi(\tu, \td,\oi) \bigg]~,\\[3mm]
\frac{\partial^2\Delta V_{O_i}}{\partial \tu \partial \td} &=& 
\frac{4\, \sdt^2}{\Delta_{O_i}}
\,\bigg[ \oi \ln \frac{m_{O_i}^4}{\tu\td} 
- (\tu -\td)\ln \frac{\tu}{\td}  
- (\oi  - \tu - \td) \Phi( \tu, \td, \oi) \bigg]~,\nn\\
\eea
where we used the shortcut
\beq
\Delta_{O_i} ~\equiv~  (\oi - \tu - \td)^2 - 4 \, \tu \td~.
\eeq
The derivatives of $\Delta V_{O_i}$ that involve $\td$ can be trivially
obtained from the ones in
eqs.~(\ref{eq:dvdstu})--(\ref{eq:dvdcdtdstu}) by means of the
replacement $\tu\!\leftrightarrow\td$, while the derivatives with
respect to all other combinations of field-dependent parameters
vanish. 

\vfill
\newpage

\bibliographystyle{utphys}
\bibliography{BGSrev}

\end{document}